\documentclass[a4paper,11pt]{article}
\usepackage[DIV12]{typearea}
\usepackage{longtable}
\usepackage{booktabs}
\usepackage{amsmath}
\usepackage{amssymb}
\usepackage{bbm}
\usepackage[utf8]{inputenc}
\usepackage{pdflscape}
\usepackage{xspace}
\usepackage[caption=false]{subfig}
\usepackage{nicefrac}
\usepackage{graphicx}
\usepackage{cite}  
\usepackage{nccfoots}
\usepackage{cancel}
\usepackage[small]{caption2}
\usepackage{multirow}

\newcommand{\rep}[1]{\ensuremath\boldsymbol{#1}}
\newcommand{\crep}[1]{\ensuremath\bar{\boldsymbol{#1}}}
\newcommand{\x}{\ensuremath\times}
\newcommand{\Z}[1]{\ensuremath{\mathbbm{Z}_{#1}}} 
\newcommand{\SO}[1]{\ensuremath{\mathrm{SO}(#1)}}
\newcommand{\SU}[1]{\ensuremath{\mathrm{SU}(#1)}}

\newcommand{\E}[1]{\ensuremath{\mathrm{E}_{#1}}}
\newcommand{\I}{\mathrm{i}}
\newcommand{\Id}{\mathbbm{1}}

\newcommand{\CP}{\ensuremath{\mathcal{CP}}\xspace}
\newcommand{\diag}{\ensuremath{\textrm{diag\,}}}

\newcommand{\nphantom}[1]{\sbox0{#1}\hspace{-\the\wd0}}

\usepackage{xcolor}

\advance \headheight by 3.0truept       
\usepackage[pdftex]{hyperref}

\hypersetup{
    pdftitle = {Lessons from eclectic flavor symmetries},
    pdfauthor = {Nilles, Ramos-Sanchez, Vaudrevange}
}
 
\begin{document}

\begin{titlepage}

\begin{flushright}
\normalsize{TUM-HEP 1256/20}
\end{flushright}

\vspace*{1.0cm}

\begin{center}
{\Large\textbf{Lessons from eclectic flavor symmetries\unboldmath}}

\vspace{1cm}

\textbf{%
Hans Peter Nilles$^{a}$, Sa\'ul Ramos--S\'anchez$^{b,c}$, Patrick K.S. Vaudrevange$^{c}$
}
\\[8mm]
\textit{$^a$\small Bethe Center for Theoretical Physics and Physikalisches Institut der Universit\"at Bonn,\\ Nussallee 12, 53115 Bonn, Germany}
\\[2mm]
\textit{$^b$\small Instituto de F\'isica, Universidad Nacional Aut\'onoma de M\'exico,\\ POB 20-364, Cd.Mx. 01000, M\'exico}
\\[2mm]
\textit{$^c$\small Physik Department T75, Technische Universit\"at M\"unchen,\\ James-Franck-Stra\ss e 1, 85748 Garching, Germany}
\end{center}

\vspace*{1cm}

\begin{abstract}
A top-down approach to the flavor problem motivated from string theory leads to the concept of 
eclectic flavor groups that combine traditional and modular flavor symmetries. To make contact with 
models constructed in the bottom-up approach, we analyze a specific example based on the eclectic 
flavor group $\Omega(1)$ (a nontrivial combination of the traditional flavor group $\Delta(54)$ and 
the finite modular group $T'$) in order to extract general lessons from the eclectic scheme.
We observe that this scheme is highly predictive since it severely restricts the possible
group representations and modular weights of matter fields. Thereby, it controls the structure
of the K\"ahler potential and the superpotential, which we discuss explicitly.
In particular, both K\"ahler potential and superpotential are shown to transform nontrivially, 
but combine to an invariant action. Finally, we find that discrete $R$-symmetries are intrinsic 
to eclectic flavor groups.
\end{abstract}

\end{titlepage}

\newpage

\section{Introduction}

We elaborate on a new approach to the flavor problem that combines traditional (discrete) flavor 
symmetries with modular flavor symmetries. This approach originated in top-down model building 
motivated by string theory. It has been developed in a series of 
papers~\cite{Baur:2019kwi,Baur:2019iai,Nilles:2020nnc}, culminating in the concept of {\it eclectic 
flavor groups}~\cite{Nilles:2020nnc}. The eclectic flavor group is a maximal extension of the 
traditional flavor group by (finite) discrete modular symmetries. It allows a new approach to the 
flavor problem compared to previous attempts that rely separately either on the traditional flavor 
symmetry or the modular flavor symmetry.

Although discrete flavor symmetries (traditional or modular) are natural ingredients in string 
theory, not many explicit models have been constructed yet in a top-down (TD) approach. Models with 
modular symmetries have been constructed in heterotic orbifolds, magnetized branes and intersecting 
D-brane models~\cite{Kobayashi:2016ovu,Kobayashi:2018rad,Kobayashi:2018bff,Kariyazono:2019ehj}. In 
particular, several promising models have been found with different orbifold 
geometries~\cite{Kobayashi:2004ya,Lebedev:2007hv,Kim:2007mt,Nilles:2008gq,Blaszczyk:2009in,Pena:2012ki,Nibbelink:2013lua,Carballo-Perez:2016ooy,Olguin-Trejo:2018wpw}. 
Even in the absence of a large number of explicit and fully satisfactory models, we think that it 
is time to combine the TD-approach with existing bottom-up (BU) models that exhibit successful fits 
to masses and mixing angles of quarks and leptons. Our analysis will clarify several conceptional 
and technical considerations that have not yet been fully addressed in the available literature, 
such as the need for the consideration of the eclectic extension and a new link between 
representations and modular weights. To illustrate these questions, we shall use a scheme based on 
the $\mathbb{T}^2/\Z3$ orbifold which appears, for example, in models based on the 
$\mathbb{T}^6/\Z3\x\Z3$ orbifold discussed in ref.~\cite{Carballo-Perez:2016ooy}. It exhibits the 
traditional flavor symmetry $\Delta(54)$, the finite modular flavor group $T' \cong [24,3]$ and the 
resulting eclectic flavor group $\Omega(1) \cong [648,533]$ (according to the classification of the 
computer program GAP~\cite{GAP4}, where the first number gives the order of the group).

\enlargethispage{0.6cm}
There is still a gap between available TD and BU constructions~\cite{
Altarelli:2005yx,            
deAdelhartToorop:2011re,     
Feruglio:2017spp             
} and there are some questions to be addressed when one tries to explicitly combine them. In BU 
constructions one freely assumes a certain modular flavor group (like $\Gamma_N \cong S_3, A_4, S_4, A_5$) 
as well as all the nontrivial modular weights and representations of these groups (like triplets 
and nontrivial singlets) that are needed to provide a successful fit to the data~\cite{
Kobayashi:2018vbk,           
Penedo:2018nmg,              
Criado:2018thu,              
Kobayashi:2018scp,           
Novichkov:2018ovf,           
Novichkov:2018nkm,           
deAnda:2018ecu,              
Okada:2018yrn,               
Kobayashi:2018wkl,           
Novichkov:2018yse,           
Ding:2019xna,                
Nomura:2019jxj,              
Novichkov:2019sqv,           
deMedeirosVarzielas:2019cyj, 
Liu:2019khw,                 
Okada:2019xqk,               
Kobayashi:2019mna,           
Ding:2019zxk,                
King:2019vhv,                
Nomura:2019lnr,              
Criado:2019tzk,              
Kobayashi:2019xvz,           
Asaka:2019vev,               
Chen:2019ewa,                
Gui-JunDing:2019wap,         
Zhang:2019ngf,               
Wang:2019ovr,                
Kobayashi:2019uyt,           
Nomura:2019xsb,              
Kobayashi:2019gtp,           
Lu:2019vgm,                  
Wang:2019xbo,                
Okada:2020dmb,               
Ding:2020yen                 
} following the influential work of Feruglio~\cite{Feruglio:2017spp}. In the cases discussed so far 
there does not yet exist a TD-model that matches all these ingredients (in particular the 
appearance of all the nontrivial representations).\\[1mm]
\indent Our TD example based on the eclectic flavor group $\Omega(1)$ is the one that comes closest to 
it. This model is suitable to illustrate the following lessons learned from the TD perspective:
\begin{itemize}
\item[i)] the representations and modular weights of the fields that appear in the low energy 
effective field theory are highly constrained,
\item[ii)] the eclectic flavor group is more predictive than the traditional flavor group or the 
finite modular group alone: it severely restricts the superpotential and the K\"ahler potential, 
\item[iii)] discrete $R$-symmetries are naturally related to the eclectic flavor group.
\end{itemize}
Once these lessons are taken into account, a meaningful link between TU and BU models can be 
discussed.

The paper is structured as follows: in section~\ref{sec:spectrum} we shall present the $\Omega(1)$ 
model in detail and identify the modular weights and representations of the fields that appear 
in the massless sector of explicit MSSM-like string models. We emphasize the possibility of having 
fields with fractional modular weights and discuss how modular weights affect the traditional flavor 
symmetry. The results are summarized in table~\ref{tab:Representations}. Section~\ref{sec:action} 
is devoted to the discussion of the effective action of the $\mathbb{T}^2/\Z3$ orbifold sector, 
including the superpotential and the K\"ahler potential.\footnote{The relevance of the K\"ahler 
potential has typically not been discussed in the existing literature of BU constructions, but has 
been emphasized in ref.~\cite{Chen:2019ewa}.} Both of them transform nontrivially under the modular 
transformation (but combine to a modular invariant action). We shall separately discuss the 
restrictions based on $T'$ and $\Delta(54)$, and illustrate the relevance of both for the eclectic 
picture. Finally, conclusions and outlook will be given in section~\ref{sec:conclusions}.

\section{Spectrum and symmetries}
\label{sec:spectrum}

We focus on symmetric Abelian toroidal orbifold compactifications of the heterotic 
string~\cite{Dixon:1985jw,Dixon:1986jc,Ibanez:1986tp} that yield both, a $T'$ finite modular 
symmetry and a $\Delta(54)$ traditional flavor symmetry. As derived in 
refs.~\cite{Kobayashi:2006wq,Nilles:2012cy}, a $\Delta(54)$ traditional flavor symmetry appears in 
compactifications endowed with a $\mathbbm{T}^2/\Z{3}$ orbifold sector with trivial Wilson line 
background fields. Moreover, such a $\mathbbm{T}^2/\Z{3}$ orbifold sector yields a finite modular 
symmetry $T' \cong\mathrm{SL}(2,3)$~\cite{Lauer:1989ax,Lerche:1989cs,Lauer:1990tm}. Importantly, 
these modular and traditional flavor symmetries do not commute and, hence, combine nontrivially to 
the so-called eclectic flavor group, $\Omega(1) \cong [648, 533]$ in this particular case, 
as explained in ref.~\cite{Nilles:2020nnc}. See also ref.~\cite{Yao:2015dwa,King:2016pgv} for BU 
flavor model building based on $\Omega(1)$, and ref.~\cite{Jurciukonis:2017mjp} for notation. 
Examples of six-dimensional orbifolds with such a 
$\mathbbm{T}^2/\Z{3}$ sector include orbifolds like $\mathbbm{T}^6/\Z6$-II, $\mathbbm{T}^6/\Z3\x\Z3$ 
and $\mathbbm{T}^6/\Z3\x\Z6$. These orbifolds are known to reproduce some properties of the MSSM 
when used to compactify the $\E{8}\times\E{8}$ heterotic 
string~\cite{Nilles:2014owa,Carballo-Perez:2016ooy,Olguin-Trejo:2018wpw,Parr:2019bta,Parr:2020oar}. 

Since the relevant flavor symmetries are fully determined by the two-dimensional \Z3 orbifold sector, 
we can restrict our discussion to this sector. There, the orbifold action is generated by 
a twist $\theta=\exp(\nicefrac{2\pi\I}{3})$ using complex coordinates for the torus $\mathbbm{T}^2$. This 
twist defines a \Z3 point group with elements $\{\Id,\theta,\theta^2\}$. Closed strings 
on $\mathbbm{T}^2/\Z{3}$ fall into three categories:\\
(i)   {\it Untwisted strings} that are trivially closed, even in uncompactified space,
associated with the element $\Id$ of the point group.\\
(ii) {\it Untwisted winding strings} that are also associated with the element $\Id$ of the point 
group but wind around some torus-directions $e_1$, $e_2$ of the orbifold. In the model discussed 
here, the winding modes are typically heavy and  therefore not relevant for our analysis.\\
(iii)  {\it Twisted strings}, which are closed only due to the action of the twist $\theta$ or $\theta^2$.

First of all, in the untwisted sector we find the K\"ahler modulus $T$ of the $\mathbbm{T}^2/\Z{3}$ 
orbifold sector that arises from the metric and the antisymmetric $B$-field of the two-torus 
$\mathbbm{T}^2$. In contrast, the complex structure modulus $U$ is fixed to 
$U=\exp(\nicefrac{2\pi\I}{3})$ for a $\mathbbm{T}^2/\Z{3}$, as is well-known. In addition, there are 
massless untwisted matter strings in four dimensions that originate from ten-dimensional gauge 
bosons $A^M$, $M=0,\ldots, 9$ of $\E{8}\times\E{8}$ (or $\SO{32}$). Depending on the internal 
vector index $M$, we denote the corresponding untwisted (i.e.\ bulk) matter fields by
\begin{equation}\label{eq:BulkMatter}
\Phi_{\text{\tiny $-1$}} \;\;\;\mathrm{if}\;\;\; M ~=~ 4,5 \qquad\mathrm{and}\qquad 
\Phi_{\text{\tiny 0}} \;\;\;\mathrm{if}\;\;\; M ~=~ 6,7,8,9\;,
\end{equation}
assuming that the $\mathbbm{T}^2/\Z{3}$ orbifold sector lies in the compactified directions $M = 4,5$. 
Note that, as discussed later in section~\ref{subsec:T'Repr}, the label $n$ of a matter field 
$\Phi_n$ gives the so-called modular weight under a finite modular transformation.

\begin{figure*}[t]
\centering{\includegraphics[width=0.55\linewidth]{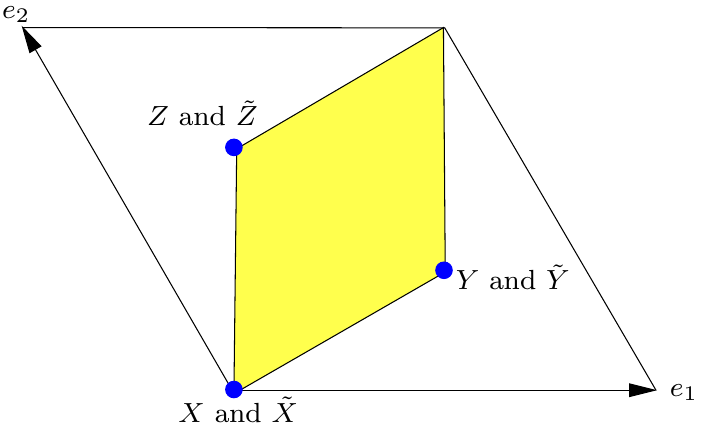}}
\vspace{-0.3cm}
\caption{The $\mathbbm{T}^2/\Z{3}$ orbifold sector: the vectors $e_1$ and $e_2$ define the 
two-torus $\mathbbm{T}^2$ that exhibits a $\Z{3}$ rotational symmetry. The fundamental domain of 
the $\mathbbm{T}^2/\Z{3}$ orbifold is depicted as the (yellow) colored region and the three inequivalent fixed points 
are represented by the (blue) bullets. $(X,Y,Z)^\mathrm{T}$ and $(\tilde{X},\tilde{Y},\tilde{Z})^\mathrm{T}$ denote localized 
triplets of matter fields corresponding to twisted strings from the $\theta$ twisted sector without 
and with oscillator excitations, respectively.}
\label{fig:Z3twistedstates}
\end{figure*}

The $\mathbbm{T}^2/\Z{3}$ orbifold sector has three fixed points, as illustrated in 
figure~\ref{fig:Z3twistedstates}. At these fixed points, additional massless strings from the 
$\theta$ and $\theta^2$ twisted sectors can be localized. For each twisted sector, there are two 
classes of massless twisted strings: either with or without oscillator excitations. Consequently, 
we have two kinds of twisted (i.e.\ localized) matter fields in the $\theta$ twisted sector. We 
denote them by
\begin{subequations}\label{eq:ThetaTwistedMatter}
\begin{eqnarray}
\Phi_{\nicefrac{-2}{3}} & = & (X, Y, Z)^\mathrm{T}                         \quad\mathrm{without\ oscillator\ excitations}\;,\\
\Phi_{\nicefrac{-5}{3}} & = & (\tilde{X}, \tilde{Y}, \tilde{Z})^\mathrm{T} \quad\mathrm{with\ one\ holomorphic\ oscillator\ excitation}\;,
\end{eqnarray}
\end{subequations}
respectively, where for example the three matter fields $X$, $Y$ and $Z$ are localized at the 
three fixed points of the $\mathbbm{T}^2/\Z{3}$ orbifold sector. We focus in this paper on 
the couplings of untwisted and $\theta$-twisted matter fields $\Phi_{\text{\tiny 0}}$, 
$\Phi_{\text{\tiny $-1$}}$, $\Phi_{\nicefrac{-2}{3}}$ and $\Phi_{\nicefrac{-5}{3}}$ only. 
For completeness, let us mention the possible massless anti-triplets of $\theta^2$-twisted matter 
fields, being
\begin{subequations}\label{eq:Theta2TwistedMatter}
\begin{eqnarray}
& \Phi_{\nicefrac{-1}{3}} & \quad\mathrm{without\ oscillator\ excitations}\;,\\
& \Phi_{\nicefrac{+2}{3}} & \quad\mathrm{with\ one\ anti}\text{-}\mathrm{holomorphic\ oscillator\ excitation}\;.
\end{eqnarray}
\end{subequations}
In general, twisted matter fields with further modular weights are possible, but we find that they 
do not appear in MSSM-like heterotic orbifold compactifications with a $\mathbbm T^2/\Z3$ sector 
possibly due to constraints similar to those presented in ref.~\cite[table 3]{Ibanez:1992hc}. As a 
remark, the CPT-partners of the $\theta^k$-twisted string states originate from the $\theta^{2k}$ 
twisted sector for $k=1,2$.

\begin{table}[t!]
\center
\begin{tabular}{|c||c|c||c|c|c|c||c|c|c|c|}
\hline
\multirow{3}{*}{sector} &\!\!matter\!\!&\multirow{3}{*}{\!\!osc.\!\!}& \multicolumn{8}{c|}{eclectic flavor group $\Omega(1)$}\\
                        &fields        &                             & \multicolumn{4}{c||}{modular $T'$ subgroup} & \multicolumn{4}{c|}{traditional $\Delta(54)$ subgroup}\\
                        &$\Phi_n$      &                             & \!\!irrep $\rep{s}$\!\! & $\rho_{\rep{s}}(\mathrm{S})$ & $\rho_{\rep{s}}(\mathrm{T})$ & $n$ & \!\!irrep $\rep{r}$\!\! & $\rho_{\rep{r}}(\mathrm{A})$ & $\rho_{\rep{r}}(\mathrm{B})$ & $\rho_{\rep{r}}(\mathrm{C})$ \\
\hline
\hline
bulk      & $\Phi_{\text{\tiny 0}}$   & no & $\rep1$             & $1$                   & $1$                   & $0$               & $\rep1$   & $1$               & $1$                   & $+1$               \\
          & $\Phi_{\text{\tiny $-1$}}$& no & $\rep1$             & $1$                   & $1$                   & $-1$              & $\rep1'$  & $1$               & $1$                   & $-1$               \\
\hline
$\theta$  & $\Phi_{\nicefrac{-2}{3}}$ & no & $\rep2'\oplus\rep1$ & $\rho(\mathrm{S})$    & $\rho(\mathrm{T})$    & $\nicefrac{-2}{3}$& $\rep3_2$ & $\rho(\mathrm{A})$& $\rho(\mathrm{B})$    & $-\rho(\mathrm{C})$\\
          & $\Phi_{\nicefrac{-5}{3}}$ & yes& $\rep2'\oplus\rep1$ & $\rho(\mathrm{S})$    & $\rho(\mathrm{T})$    & $\nicefrac{-5}{3}$& $\rep3_1$ & $\rho(\mathrm{A})$& $\rho(\mathrm{B})$    & $+\rho(\mathrm{C})$\\
\hline
$\theta^2$& $\Phi_{\nicefrac{-1}{3}}$ & no & $\rep2''\oplus\rep1$& $(\rho(\mathrm{S}))^*$& $(\rho(\mathrm{T}))^*$& $\nicefrac{-1}{3}$& $\crep3_1$& $\rho(\mathrm{A})$& $(\rho(\mathrm{B}))^*$& $+\rho(\mathrm{C})$\\
          & $\Phi_{\nicefrac{+2}{3}}$ & yes& $\rep2''\oplus\rep1$& $(\rho(\mathrm{S}))^*$& $(\rho(\mathrm{T}))^*$& $\nicefrac{+2}{3}$& $\crep3_2$& $\rho(\mathrm{A})$& $(\rho(\mathrm{B}))^*$& $-\rho(\mathrm{C})$\\
\hline
\hline
super-    & \multirow{2}{*}{$\mathcal{W}$} & \multirow{2}{*}{-} & \multirow{2}{*}{$\rep1$} & \multirow{2}{*}{$1$} & \multirow{2}{*}{$1$} & \multirow{2}{*}{$-1$} & \multirow{2}{*}{$\rep1'$} & \multirow{2}{*}{$1$} & \multirow{2}{*}{$1$} & \multirow{2}{*}{$-1$} \\
\!\!potential\!\! & & & & & & & & & & \\
\hline
\end{tabular}
\caption{$T'$ and $\Delta(54)$ irreducible representations of (massless) matter fields $\Phi_n$ 
with modular weights $n$ in MSSM-like heterotic orbifold compactifications with a $\mathbbm{T}^2/\Z3$ 
sector, see refs.~\cite{Baur:2019kwi,Baur:2019iai} for the derivations. $T'$ and $\Delta(54)$ combine 
nontrivially to the $\Omega(1) \cong [648, 533]$ eclectic flavor group~\cite{Nilles:2020nnc}, 
generated by $\rho_{\rep{s}}(\mathrm{S})$, $\rho_{\rep{s}}(\mathrm{T})$, 
$\rho_{\rep{r}}(\mathrm{A})$ and $\rho_{\rep{r}}(\mathrm{B})$. For $\rho_{\rep{r}}(\mathrm{C})$, 
both $\mathrm{C}=\mathrm{S}^2$ and the modular weight $n$ are important, as discussed later in 
eq.~\eqref{eq:Delta54GeneratorC}. Untwisted matter fields $\Phi_n$ (with integer modular weights 
$n$) form one-dimensional representations, while twisted matter fields $\Phi_n$ (with fractional 
modular weights $n$) form (anti-)triplet representations.}
\label{tab:Representations}
\end{table}

\subsection[$T'$ representations]{\boldmath $T'$ representations \unboldmath}
\label{subsec:T'Repr}

Let us discuss the modular transformation properties of untwisted and twisted matter fields 
$\Phi_n$ for orbifolds having a $\mathbbm{T}^2/\Z{3}$ sector.

The modular group $\mathrm{SL}(2,\Z{})$ is defined as
\begin{equation}\label{eq:gamma}
\gamma ~=~ \left(\begin{array}{cc} a & b \\ c & d \end{array}\right) ~\in~ \mathrm{SL}(2,\Z{}) \quad\Leftrightarrow\quad ad-bc ~=~ 1\,\quad\text{ and }\quad a,b,c,d~\in~\Z{}\,.
\end{equation}
It can be generated by two elements,
\begin{equation}\label{eq:SL2ZGenerators}
\mathrm{S} ~:=~ \left(\begin{array}{cc} 0 & 1 \\ -1 & 0 \end{array}\right) \quad\mathrm{and}\quad 
\mathrm{T} ~:=~ \left(\begin{array}{cc} 1 & 1 \\ 0 & 1 \end{array}\right)\;,\qquad \mathrm{S}, \mathrm{T} ~\in~ \mathrm{SL}(2,\Z{})\;,
\end{equation}
which satisfy the defining relations
\begin{equation}\label{eq:SL2Z}
\mathrm{S}^4 ~=~ (\mathrm{S}\,\mathrm{T})^3 ~=~ \Id \quad\mathrm{and}\quad \mathrm{S}^2\,\mathrm{T} ~=~ \mathrm{T}\,\mathrm{S}^2
\end{equation}
of $\mathrm{SL}(2,\Z{})$. Under a general modular transformation $\gamma\in\mathrm{SL}(2,\Z{})$ 
from eq.~\eqref{eq:gamma}, the K\"ahler modulus $T$ transforms as
\begin{equation}\label{eq:TrafoWithGamma}
T ~\stackrel{\gamma}{\longrightarrow}~ \frac{a\,T+b}{c\,T+d}\;.
\end{equation}
Since $T$ transforms identically for $\pm\gamma$, it feels only $\mathrm{PSL}(2,\Z{})$ instead of 
the full $\mathrm{SL}(2,\Z{})$ modular group. In contrast, a general matter field $\Phi_n$ 
transforms under $\gamma \in \mathrm{SL}(2,\Z{})$ as
\begin{equation}\label{eq:ModularTransformationOfPhi}
\Phi_n ~\stackrel{\gamma}{\longrightarrow}~ {\Phi_n}' ~:=~ (c\,T+d)^n\, \rho_{\rep{s}}(\gamma)\,\Phi_n\;,
\end{equation}
where $(c\,T + d)^n$ is the so-called automorphy factor with modular weight $n\in\mathbbm{Q}$. 
Note that fractional modular weights of both signs are common to string theory, see for example 
refs.~\cite{Ferrara:1989qb,Ibanez:1992hc}. Moreover, for orbifolds with a $\mathbbm{T}^2/\Z{3}$ 
sector, the matrices $\rho_{\rep{s}}(\gamma)$ build a (reducible or irreducible) representation 
$\rep{s}$ of the finite modular group $T'\cong\mathrm{SL}(2,3)$, which satisfy the defining 
relations of $T'$,
\begin{equation}\label{eq:TprimePresentation}
\rho_{\rep{s}}(\mathrm{S})^4 ~=~ \rho_{\rep{s}}(\mathrm{T})^3 ~=~ (\rho_{\rep{s}}(\mathrm{S})\,\rho_{\rep{s}}(\mathrm{T}))^3 ~=~ \Id\;, \quad (\rho_{\rep{s}}(\mathrm{S}))^2 \rho_{\rep{s}}(\mathrm{T})~=~ \rho_{\rep{s}}(\mathrm{T}) (\rho_{\rep{s}}(\mathrm{S}))^2\;,
\end{equation}
cf. eq.~\eqref{eq:SL2Z}. In more detail, for the generators $\mathrm{S}$ and $\mathrm{T}$ of 
$\mathrm{SL}(2,\Z{})$, given in eq.~\eqref{eq:SL2ZGenerators}, a general matter field $\Phi_n$ 
transforms as
\begin{subequations}\label{eq:ModularTrafoOfFields}
\begin{eqnarray}
\Phi_n &\stackrel{\mathrm{S}}{\longrightarrow}& {\Phi_n}' ~:=~ (-T)^n\, \rho_{\rep{s}}(\mathrm{S})\,\Phi_n\;,\label{eq:ModularSTrafoOfFields}\\
\Phi_n &\stackrel{\mathrm{T}}{\longrightarrow}& {\Phi_n}' ~:=~          \rho_{\rep{s}}(\mathrm{T})\,\Phi_n\;.
\end{eqnarray}
\end{subequations}
In the following, we specify $\rho_{\rep{s}}(\mathrm{S})$ and $\rho_{\rep{s}}(\mathrm{T})$ for the 
matter fields $\Phi_n$ of our orbifold theory: In the untwisted sector, there are two kinds of bulk 
fields, denoted by $\Phi_{\text{\tiny 0}}$ and $\Phi_{\text{\tiny $-1$}}$ with modular weight $n=0$ 
and $n=-1$, respectively, see eq.~\eqref{eq:BulkMatter}. Both transform as trivial singlets of 
$T'$, i.e.\ $\rho_{\rep{1}}(\mathrm{S})=\rho_{\rep{1}}(\mathrm{T})=1$. In the twisted sectors of 
the orbifold, where matter fields build triplets associated to the three fixed points of the 
$\mathbbm{T}^2/\Z{3}$ orbifold sector, we have to distinguish between four cases: matter fields 
$\Phi_n$ from the $\theta$ or $\theta^2$ twisted sector with or without oscillator excitations, see 
eqs.~\eqref{eq:ThetaTwistedMatter} and~\eqref{eq:Theta2TwistedMatter}. They carry different modular 
weights $n$ and transform in different three-dimensional representations $\rep{s}$ of $T'$, 
as displayed in 
table~\ref{tab:Representations}. In all four cases, $\rho_{\rep{s}}(\mathrm{S})$ 
and $\rho_{\rep{s}}(\mathrm{T})$ are related to the $3 \times 3$ matrices
\begin{equation}\label{eq:ModularTrafoOfTwistedStringsRep}
\rho(\mathrm{S}) ~:=~ \frac{\I}{\sqrt{3}}\left(\begin{array}{ccc} 1 & 1 & 1 \\ 1 & \omega^2 &\omega \\ 1 & \omega & \omega^2 \end{array}\right) \qquad\mathrm{and}\qquad
\rho(\mathrm{T}) ~:=~ \left(\begin{array}{ccc} \omega^2 & 0 & 0 \\ 0 & 1 &0 \\ 0 & 0 & 1 \end{array}\right)\;,
\end{equation}
where $\omega:=\exp(\nicefrac{2\pi\I}{3})$. Note that we use a different convention compared to 
ref.~\cite{Baur:2019iai}: we redefine $\mathrm{S}^3$ from ref.~\cite{Baur:2019iai} to $\mathrm{S}$. 
Consequently, we are now using the presentation eq.~\eqref{eq:SL2Z} of $\mathrm{SL}(2,\Z{})$ 
instead of $\mathrm{S}^4 = \Id$, $\mathrm{S}^2 = (\mathrm{S}\,\mathrm{T})^3$. For this change of 
convention, we redefine the outer automorphism $\hat{K}_{\mathrm{S}}$ of the Narain lattice (as 
defined in ref.~\cite{Baur:2019iai}) from $\hat{K}_{\mathrm{S}}^3$ to $\hat{K}_{\mathrm{S}}$ (and 
analogously for $\hat{C}_{\mathrm{S}}$). This results in a redefinition of $\rho(\mathrm{S})^3$ to 
$\rho(\mathrm{S})$.

The three-dimensional $T'$ representations $\rep{s}$ of twisted matter fields (listed in 
table~\ref{tab:Representations}) are reducible representations. They decompose into irreducible 
representations as doublets plus trivial singlets of $T'$. In more detail, for the triplet 
$\Phi_{\nicefrac{-2}{3}}=(X,Y,Z)^\mathrm{T}$ of $\theta$-twisted fields without oscillator 
excitations we find the decomposition $\rep{2}'\oplus\rep{1}$ using the $T'$ conventions of 
ref.~\cite{Ishimori:2010au} with $p=\I$. Explicitly, the doublet $\rep{2}'$ and the singlet 
$\rep{1}$ are given by the linear combinations
\begin{equation}
  \rep2':~ \left( \begin{array}{c} \frac{1}{\sqrt2}(Y+Z) \\ -X \end{array}\right)\qquad\text{and}\qquad
  \rep1:~ \frac{1}{\sqrt2}(Y-Z)\,.
\end{equation}
An analogous combination holds for the $\theta$-twisted fields with oscillator excitations 
$\Phi_{\nicefrac{-5}{3}}$.

For the anti-triplet $\Phi_{\nicefrac{-1}{3}} = (U,V,W)^\mathrm{T}$ of twisted fields from the 
$\theta^2$ twisted sector without oscillator excitations, the following linear combinations build 
the doublet $\rep2''$ and the trivial singlet $\rep{1}$ of $T'$
\begin{equation}
  \rep2'':~ \left( \begin{array}{c} U \\ \frac{1}{\sqrt2}(V+W) \end{array}\right)\qquad\text{and}\qquad
  \rep1:~ \frac{1}{\sqrt2}(V-W)\,.
\end{equation}
An analogous combination holds for the $\theta^2$-twisted fields with oscillator excitations 
$\Phi_{\nicefrac{+2}{3}}$.

\subsection[Delta(54) representations]{\boldmath $\Delta(54)$ representations \unboldmath}
\label{subsec:Delta54Repr}

In addition to a $T'$ finite modular symmetry, our $\mathbbm{T}^2/\Z{3}$ orbifold sector enjoys a 
$\Delta(54)$ traditional flavor symmetry~\cite{Kobayashi:2006wq}. $\Delta(54)$ can be generated by 
three elements, denoted by $\mathrm{A}$, $\mathrm{B}$ and $\mathrm{C}$. From a string point of view 
(based on the Narain space group~\cite{GrootNibbelink:2017usl} and its outer automorphisms), the 
generators $\mathrm{A}$ and $\mathrm{B}$ originate from translations, while 
$\mathrm{C}$ is given by a $180^\circ$ rotation~\cite{Baur:2019kwi,Baur:2019iai}. The different 
origin of $\mathrm{A}$ and $\mathrm{B}$ as translations on one side and $\mathrm{C}$ as a 
rotation on the other has important consequences, as we discuss in the following.

To do so, let us describe how $\Delta(54)$ acts on matter fields. Take a generator 
$\mathrm{g}\in\{\mathrm{A},\mathrm{B},\mathrm{C}\}$ of $\Delta(54)$. Then, for matter fields 
originating from the orbifold bulk, we find
\begin{subequations}\label{eq:Delta54TrafoOfUntwistedStrings}
\begin{eqnarray}
\Phi_{\text{\tiny 0}}    &\stackrel{\mathrm{g}}{\longrightarrow}& {\Phi_{\text{\tiny 0}}}'    ~=~ \Phi_{\text{\tiny 0}}\;,\\
\Phi_{\text{\tiny $-1$}} &\stackrel{\mathrm{g}}{\longrightarrow}& {\Phi_{\text{\tiny $-1$}}}' ~=~ \rho_{\rep{1}'}\!(\mathrm{g})\,\Phi_{\text{\tiny $-1$}}\;.
\end{eqnarray}
\end{subequations}
Moreover, $g$ acts on triplets of localized matter fields $\Phi_n$ from the $\theta$ twisted sector 
as\footnote{In this work, $\Delta(54)$ triplets are denoted by $\rep3_1$, $\rep3_2$, $\crep3_1$ and 
$\crep3_2$ and correspond, in the conventions of ref.~\cite{Ishimori:2010au}, to $\rep{3}_{1(1)}$, 
$\rep{3}_{2(1)}$, $\rep{3}_{1(2)}$ and $\rep{3}_{2(2)}$, respectively.}
\begin{subequations}\label{eq:Delta54TrafoOfThetaTwistedStrings}
\begin{eqnarray}
\Phi_{\nicefrac{-2}{3}} &\stackrel{\mathrm{g}}{\longrightarrow}& {\Phi_{\nicefrac{-2}{3}}}' ~=~ \rho_{\rep{3}_2}\!(\mathrm{g})\,\Phi_{\nicefrac{-2}{3}}\;,\\
\Phi_{\nicefrac{-5}{3}} &\stackrel{\mathrm{g}}{\longrightarrow}& {\Phi_{\nicefrac{-5}{3}}}' ~=~ \rho_{\rep{3}_1}\!(\mathrm{g})\,\Phi_{\nicefrac{-5}{3}}\;,
\end{eqnarray}
\end{subequations}
while for twisted fields from the $\theta^2$ twisted sectors we have
\begin{subequations}\label{eq:Delta54TrafoOfTheta2TwistedStrings}
\begin{eqnarray}
\Phi_{\nicefrac{-1}{3}} &\stackrel{\mathrm{g}}{\longrightarrow}& {\Phi_{\nicefrac{-1}{3}}}' ~=~ \rho_{\crep{3}_1}\!(\mathrm{g})\,\Phi_{\nicefrac{-1}{3}}\;,\\
\Phi_{\nicefrac{+2}{3}} &\stackrel{\mathrm{g}}{\longrightarrow}& {\Phi_{\nicefrac{+2}{3}}}' ~=~ \rho_{\crep{3}_2}\!(\mathrm{g})\,\Phi_{\nicefrac{+2}{3}}\;.
\end{eqnarray}
\end{subequations}
The corresponding three-dimensional matrix representations of $\mathrm{A}$, $\mathrm{B}$ and 
$\mathrm{C}$ are given in terms of the matrices
\begin{equation}\label{eq:Delta54Generators}
\rho(\mathrm{A}) ~:=~ \left(\begin{array}{ccc} 0 & 1 & 0 \\ 0 & 0 & 1 \\ 1 & 0 & 0 \end{array}\right)\;, \quad \rho(\mathrm{B}) ~:=~ \left(\begin{array}{ccc} 1 & 0 & 0 \\ 0 & \omega & 0 \\ 0 & 0 & \omega^2 \end{array}\right) \quad\mathrm{and}\quad \rho(\mathrm{C}) ~:=~ \left(\begin{array}{ccc} 1 & 0 & 0 \\ 0 & 0 & 1 \\ 0 & 1 & 0 \end{array}\right)\;,
\end{equation}
see table~\ref{tab:Representations}. Let us stress that the transformation property of matter 
fields $\Phi_n$ under the generator $\mathrm{C}$ depends not only on the twisted sector of $\Phi_n$ 
but also on its modular weight $n$, see eq.~\eqref{eq:Delta54TrafoOfUntwistedStrings} for fields 
form the bulk, eq.~\eqref{eq:Delta54TrafoOfThetaTwistedStrings} for $\theta$-twisted fields, and 
eq.~\eqref{eq:Delta54TrafoOfTheta2TwistedStrings} for $\theta^2$-twisted fields. 

Before we analyze 
the origin of this behavior, let us briefly comment on the $\Delta(54)$ generators $\mathrm{A}$ and 
$\mathrm{B}$.
Since $\mathrm{A}$ and $\mathrm{B}$ correspond to translations in the Narain lattice, twisted 
matter fields from the same twisted sector transform independently of oscillator excitations 
under $\mathrm{A}$ and $\mathrm{B}$. Moreover, a matter field from the $\theta^2$ twisted sector 
transforms in the complex conjugate representation compared to a matter field from the $\theta$ 
twisted sector~\cite{Baur:2019iai}. Furthermore, one can check easily that the generators 
$\mathrm{A}$ and $\mathrm{B}$ generate a $\Delta(27)\cong \Z{3}^{\mathrm{(perm.)}}\ltimes(\Z{3}^{\mathrm{(PG)}}\!\times\Z{3})$ 
subgroup of $\Delta(54)$. Here, as one sees from eq.~\eqref{eq:Delta54Generators}, the 
transformation $\mathrm{A}$ generates the $\Z{3}^{\mathrm{(perm.)}}$ subgroup of the full $S_3$ 
permutation symmetry within $\Delta(54)$~\cite{Kobayashi:2006wq}. In addition, the 
$\Z{3}^{\mathrm{(PG)}}\!\times\Z{3}$ subgroup of $\Delta(27)$ corresponds to the point and space 
group selection rules~\cite{Hamidi:1986vh,Ramos-Sanchez:2018edc} generated by
\begin{equation}\label{eq:Z3xZ3}
\mathrm{A}^2\mathrm{B}^2\mathrm{A}\,\mathrm{B} \quad\mathrm{and}\quad \mathrm{B}\;,
\end{equation}
respectively. Explicitly, for twisted matter fields from the $\theta$ twisted sector, 
eq.~\eqref{eq:Z3xZ3} yields
\begin{equation}
\label{eq:pointgroupselectionrule}
\big(\rho(\mathrm{A})\big)^2 \big(\rho(\mathrm{B})\big)^2 \rho(\mathrm{A})\, \rho(\mathrm{B}) ~=~ \diag(\omega,\omega,\omega) \quad\mathrm{and}\quad \rho(\mathrm{B}) ~=~ \diag(1,\omega,\omega^2)\;,
\end{equation}
as expected from the $\Z{3}^{\mathrm{(PG)}}\!\times\Z{3}$ point and space group selection rules. 
Analogously, one can check eq.~\eqref{eq:pointgroupselectionrule} for twisted fields from the 
$\theta^2$ twisted sector. Let us emphasize that this $\Z{3}\times\Z{3}$ is not built in by hand in 
order to identify $\Delta(54)$ as the traditional flavor symmetry of the $\mathbbm{T}^2/\Z{3}$ 
orbifold sector, as done in ref.~\cite{Kobayashi:2006wq}, but a direct consequence from translations 
in the Narain formulation of strings on orbifolds. 

Note that for each pair of matter fields in 
eqs.~\eqref{eq:Delta54TrafoOfUntwistedStrings},~\eqref{eq:Delta54TrafoOfThetaTwistedStrings} 
and~\eqref{eq:Delta54TrafoOfTheta2TwistedStrings}, the $\Delta(54)$ representations $\rep{r}$ 
depend on the respective modular weights $n$. This is due to the fact that the $\Delta(54)$ generator 
$\mathrm{C}$ is related to the modular $\mathrm{S}$ transformation via $\mathrm{C} = \mathrm{S}^2$, 
see ref.~\cite{Baur:2019iai}. Since the K\"ahler modulus $T$ is invariant under 
$\mathrm{S}^2$, the transformation $\mathrm{C}$ can be interpreted as an element of 
 the traditional flavor group. In more 
detail, applying the modular $\mathrm{S}$ transformation eq.~\eqref{eq:ModularSTrafoOfFields} twice 
for a field $\Phi_n$ that transforms in a representation $\rep{s}$ of $T'$ yields
\begin{equation}
\label{eq:S^2onPhi}
\Phi_n ~\stackrel{\mathrm{S}}{\longrightarrow}~ (-T)^n \rho_{\rep{s}}(\mathrm{S})\,\Phi_n 
       ~\stackrel{\mathrm{S}}{\longrightarrow} \left(\nicefrac{1}{T}\right)^n (-T)^n \big(\rho_{\rep{s}}(\mathrm{S})\big)^2\,\Phi_n
         ~=~ (-1)^n \big(\rho_{\rep{s}}(\mathrm{S})\big)^2\,\Phi_n\;.
\end{equation}
Consequently, the $\Delta(54)$ generator $\mathrm{C} = \mathrm{S}^2$ acts on a matter field 
$\Phi_n$ as
\begin{equation}\label{eq:Delta54GeneratorC}
\Phi_n ~\stackrel{\mathrm{C}}{\longrightarrow}~ {\Phi_n}' ~=~ \rho_{\rep{r}}(\mathrm{C})\,\Phi_n\;, \qquad\mathrm{where}\quad 
          \rho_{\rep{r}}(\mathrm{C}) ~:=~ (-1)^n \big(\rho_{\rep{s}}(\mathrm{S})\big)^2\;.
\end{equation}
Hence, $\rho_{\rep{r}}(\mathrm{C})$ is a matrix representation $\rep{r}$ of $\Delta(54)$ which 
depends on both, the modular weight $n$ and the representation matrix $\rho_{\rep{s}}(\mathrm{S})$ 
of $T'$. Consider for example the bulk matter fields $\Phi_{\text{\tiny 0}}$ and 
$\Phi_{\text{\tiny $-1$}}$: At a generic point in moduli space massless strings from the bulk must 
have vanishing winding and Kaluza-Klein numbers. Hence, $\Phi_{\text{\tiny 0}}$ and 
$\Phi_{\text{\tiny $-1$}}$ are invariant under the $\Delta(54)$ generators $\mathrm{A}$ and 
$\mathrm{B}$ and they form trivial singlets of $T'$, i.e.\ 
$\rho_{\rep{1}}(\mathrm{S})=\rho_{\rep{1}}(\mathrm{T})=1$, see refs.~\cite{Baur:2019kwi,Baur:2019iai} 
and table~\ref{tab:Representations}. Yet, due to their modular weights being $n=0$ or $n=-1$ the 
respective representations of the $\Delta(54)$ generator $\mathrm{C}$ are given by
\begin{subequations}
\begin{eqnarray}
\rho_{\rep{1}}(\mathrm{C})  & = & (-1)^0 \;\;~=~ +1  \quad\mathrm{for}\quad\Phi_{\text{\tiny 0}}\;,\\
\rho_{\rep{1}'}(\mathrm{C}) & = & (-1)^{-1}  ~=~ -1  \quad\mathrm{for}\quad\Phi_{\text{\tiny $-1$}}\;,
\end{eqnarray}
\end{subequations}
as stated already in eq.~\eqref{eq:Delta54TrafoOfUntwistedStrings} and in 
table~\ref{tab:Representations}. The analogous discussion applies to twisted matter fields from 
eqs.~\eqref{eq:Delta54TrafoOfThetaTwistedStrings} and~\eqref{eq:Delta54TrafoOfTheta2TwistedStrings}. 
Note that in these cases $(-1)^n$ is multivalued, since the modular weight $n$ is fractional. For 
example, for the representation matrix $\rho_{\rep{r}}(\mathrm{C})=(-1)^n (\rho_{\rep{s}}(\mathrm{S}))^2$ 
of the $\theta$-twisted matter fields $\Phi_{\nicefrac{-2}{3}}$ we obtain a factor
\begin{equation}
\label{eq:multivaluedfactor}
(-1)^{\nicefrac{-2}{3}} ~=~ \{1,\omega,\omega^2\}\;,
\end{equation}
while $(-1)^{\nicefrac{-5}{3}}=-(-1)^{\nicefrac{-2}{3}}$ for $\Phi_{\nicefrac{-5}{3}}$. Then, any 
of the values of $(-1)^{\nicefrac{-2}{3}}$ in the definition of $\rho_{\rep{r}}(\mathrm{C})$ in 
eq.~\eqref{eq:Delta54GeneratorC} can be absorbed by multiplying powers of the $\Z{3}^{\mathrm{(PG)}}$ 
point group generator~\eqref{eq:pointgroupselectionrule}. This implies that 
eq.~\eqref{eq:Delta54GeneratorC} reads for example for the twisted matter fields 
$\Phi_{\nicefrac{-2}{3}}$ 
\begin{equation}
\rho_{\rep{r}}(\mathrm{C}) ~=~ \big(\rho_{\rep{s}}(\mathrm{S})\big)^2 ~=~ \left(\begin{array}{ccc} -1 & 0 & 0 \\ 0 & 0 & -1 \\ 0 &-1 & 0 \end{array}\right) ~=~ -\rho(\mathrm{C})\;,
\end{equation}
up to point group elements and $\rho(\mathrm{C})$ is defined in eq.~\eqref{eq:Delta54Generators}. 
Thus, the $\theta$-twisted matter fields $\Phi_{\nicefrac{-2}{3}}$ with modular weight 
$n=\nicefrac{-2}{3}$ transform in the representation $\rep{r}=\rep{3}_2$ of $\Delta(54)$. 
Analogously, we find that for $\Phi_{\nicefrac{-5}{3}}$ the representation matrix reads 
$\rho_{\rep{r}}(\mathrm{C}) = \rho(\mathrm{C})$ and, hence, $\rep{r}=\rep{3}_1$. Note that the 
different $\Delta(54)$ representations $\rep{3}_2$ and $\rep{3}_1$ for $\theta$-twisted strings 
without and with oscillator excitation (denoted by $\Phi_{\nicefrac{-2}{3}}$ and 
$\Phi_{\nicefrac{-5}{3}}$, respectively) have an intuitive interpretation in string theory: Since 
$\mathrm{C}$ acts as a $180^\circ$ rotation in the $\mathbbm{T}^2/\Z{3}$ orbifold sector, an 
oscillator excitation picks up an additional factor $-1$ under $\mathrm{C}$, see e.g. 
ref.~\cite{Kobayashi:2004ya}. This fact gives rise to the $\Delta(54)$ representations $\rep{3}_2$ 
and $\rep{3}_1$ which differ only by a minus-sign for the generator $\mathrm{C}$.

We point out that $\Delta(54)$ doublets do not appear in the \emph{massless} spectrum of strings in 
the $\mathbbm{T}^2/\Z{3}$ orbifold sector for an arbitrary value of the K\"ahler modulus $T$. 
However, $\Delta(54)$ doublets do appear as (generically massive) winding strings which are 
instrumental for \CP violation~\cite{Nilles:2018wex}. Only at some special points in moduli space 
(e.g.\ $T = \exp(\nicefrac{2\pi\I}{3})$) some of these doublets can become massless.

\subsection[Comment on fractional modular weights]{\boldmath Comment on fractional modular weights \unboldmath}
\label{subsec:FractionalModularWeights}

Let us emphasize a remarkable connection between matter fields with fractional modular weights $n$ 
and the traditional flavor symmetry. As we have seen, the generator $\mathrm{C} = \mathrm{S}^2$ is a 
traditional symmetry as it leaves the K\"ahler modulus $T$ invariant, cf. eq.~\eqref{eq:S^2onPhi}. 
From the defining relations~\eqref{eq:TprimePresentation} of $T'$ we know that 
$(\rho_{\rep{s}}(\mathrm{S}))^4=\Id$. Hence, one might expect that $\mathrm{C}$ generates a $\Z{2}$ 
symmetry. However, due to the presence of the automorphy factor with modular weight $n$ we obtain 
form eq.~\eqref{eq:Delta54GeneratorC}
\begin{equation}\label{eq:Csquared}
\big(\rho_{\rep{r}}(\mathrm{C})\big)^2 ~=~ (-1)^{2n} \big(\rho_{\rep{s}}(\mathrm{S})\big)^4 ~=~ (-1)^{2n}\Id\;,
\end{equation}
for the transformation $\mathrm{C}^2 = \mathrm{S}^4$ of a matter field $\Phi_n$. If the modular 
weights of all fields are integer, the naive expectation is correct and $\mathrm{C} = \mathrm{S}^2$ 
generates a $\Z{2}$ traditional flavor symmetry. However, in string theory fractional modular 
weights appear frequently, for example, $n=\nicefrac{-2}{3}$ for the $\theta$-twisted matter field 
$\Phi_{\nicefrac{-2}{3}}$ in our $\mathbbm{T}^2/\Z3$ orbifold discussion. Using that 
eq.~\eqref{eq:Csquared} is multivalued for a fractional modular weight like $n=\nicefrac{-2}{3}$, 
see eq.~\eqref{eq:multivaluedfactor}, we find that $(\rho_{\rep{r}}(\mathrm{C}))^2$ gives rise to 
a nontrivial $\Z{3}$ traditional flavor symmetry, which coincides in this case with the 
$\Z{3}^{\mathrm{(PG)}}$ point group selection rule given in eq.~\eqref{eq:pointgroupselectionrule}. 

Consequently, we arrive at a general result that is also valid in bottom-up constructions: in the 
eclectic picture, consistency between the modular symmetry and the traditional flavor symmetry 
constrains the allowed choices for fractional modular weights. On the one hand, if one first 
specifies the finite modular symmetry and some fractional weights for matter fields, the traditional 
flavor symmetry has to be chosen accordingly. On the other hand, if one chooses first the 
traditional flavor symmetry and looks for its eclectic extension by a modular symmetry (without 
enlarging the traditional flavor symmetry further), the set of consistent fractional modular 
weights is limited.

\subsection{Summary}
\label{sec:spectrumsummary}

In summary, in this section we have described the transformation properties of massless matter 
fields appearing in MSSM-like models with a $\mathbbm{T}^2/\Z3$ orbifold sector under both, modular 
and traditional flavor symmetries. This sector is naturally endowed with an $\Omega(1)$ eclectic 
flavor symmetry, which comprises the $T'$ finite modular symmetry and the $\Delta(54)$ traditional 
flavor symmetry. The representations and modular weights $n$ of all six admissible types of 
massless matter fields $\Phi_n$ are determined by the compactification. Relevant details can be 
read off from table~\ref{tab:Representations}.

It should be emphasized that only a subset of $\Delta(54)$ and $T'$ representations and only a 
couple of (fractional) modular weights, which are consistent with both the modular and the 
traditional flavor symmetries, are realized among the massless states in string theory. This has 
important consequences for explicit TD model building and the connection to the BU approach.

\section[Effective action of the T2/Z3 orbifold sector]{\boldmath Effective action of the $\mathbbm{T}^2/\Z{3}$ orbifold sector\unboldmath}
\label{sec:action}

The phenomenological consequences of compactifying string theory on an orbifold arise from its 
low-energy effective field theory limit, which in our case is a theory of $\mathcal{N}=1$ 
supergravity in four dimensions. In this work, we focus on the superpotential $\mathcal{W}$ and the 
K\"ahler potential $K$ for (twisted) matter fields and construct the most general $\mathcal{W}$ and 
$K$, consistent with all symmetries of the $\mathbbm{T}^2/\Z{3}$ orbifold sector. This includes the 
traditional flavor symmetry $\Delta(54)$ that combines with the finite modular symmetry $T'$ (given 
as a realization of the full modular symmetry $\mathrm{SL}(2,\Z{})$ for twisted matter fields) to 
the eclectic flavor symmetry $\Omega(1)$. Since $\mathcal{W}$ and $K$ depend on the (dimensionless) 
K\"ahler modulus $T$ and the matter fields $\Phi_n$, the properties of $\mathcal{W}$ and $K$ must 
combine to yield a theory that is invariant under these symmetries. 

The superpotential is a holomorphic function of the matter fields $\Phi_n$, whose coefficients are 
in general modular forms $\hat Y^{(n_Y)}(T)$ (with integer modular weights $n_Y$) of the K\"ahler 
modulus $T$. Under a general modular transformation $\gamma\in\mathrm{SL}(2,\mathbbm{Z})$, the 
superpotential must transform as~\cite{Lauer:1989ax,Lerche:1989cs,Lauer:1990tm}
\begin{equation}\label{eq:WgammaTrafo}
  \mathcal{W}(T,\Phi_n)  ~\stackrel{\gamma}{\longrightarrow}~  \mathcal{W}\left(\frac{a\,T+b}{c\,T+d},\Phi_n'\right) 
                       ~=~ (c\,T+d)^{-1}\, \mathcal{W}(T,\Phi_n)\;,
\end{equation}
where the transformed matter fields ${\Phi_n}'$ are given in eq.~\eqref{eq:ModularTransformationOfPhi}. 
Thus, the superpotential behaves like a chiral superfield with modular weight $n=-1$, as we will 
discuss in more detail later in eq.~\eqref{eq:Sl2ZPlusKahler}. This implies in particular that 
under $\mathrm{C}=\mathrm{S}^2$ (which leaves the modulus $T$ invariant) the superpotential 
transforms as
\begin{equation}\label{eq:WS2Trafo}
  \mathcal{W}(T,\Phi_n) ~\stackrel{\mathrm{S}^2}{\longrightarrow}~ -\mathcal{W}(T,\Phi_n)\;,
\end{equation}
using the automorphy factor $(0\cdot T -1)^{-1}=-1$ for $\mathrm{S}^2 = -\Id$, see 
eq.~\eqref{eq:SL2ZGenerators}. Hence, $\mathrm{C}=\mathrm{S}^2$ acts as an $R$-symmetry that 
transforms the Grassmann number $\vartheta$ of $\mathcal{N}=1$ superspace as $\vartheta \rightarrow \I\,\vartheta$ 
such that $\mathcal{L}\supset\int\!\mathrm{d}^2\vartheta\, \mathcal{W}$ is invariant. This might 
have been expected since $\mathrm{C}$ is defined as a $180^\circ$ rotation in the $\mathbbm{T}^2/\Z{3}$ 
orbifold sector~\cite{Baur:2019kwi,Baur:2019iai}. Moreover, $\mathrm C$ acts as a \Z2 
$R$-symmetry on bosons but as a \Z4 $R$-symmetry on fermions. In this sense, $\Delta(54)$ is the 
traditional flavor symmetry of the bosonic particle content. 

Furthermore, under the generators $\mathrm{A}$ and $\mathrm{B}$ of the traditional flavor group 
$\Delta(54)$ the superpotential must be invariant, i.e.
\begin{equation}
\mathcal{W}(T,\Phi_n) ~\stackrel{\mathrm{A}, \mathrm{B}}{\longrightarrow}~ \mathcal{W}(T,{\Phi_n}') ~=~ \mathcal{W}(T,\Phi_n)\;,
\end{equation}
using that the modulus $T$ is invariant under $\mathrm{A}$ and $\mathrm{B}$. In summary, the 
transformations under $\mathrm{A}$, $\mathrm{B}$, and $\mathrm{C}$ imply that $\mathcal{W}$ builds 
a $\rep1'$ representation of $\Delta(54)$.

Let us stress two important results concerning the $R$-symmetry transformation eq.~\eqref{eq:WS2Trafo}:
\begin{enumerate}
\item[i)] First, this $R$-symmetry is part of both, modular and traditional flavor transformations: 
$\mathrm{S}^2\in\mathrm{SL}(2,\mathbbm{Z})$ and $\mathrm{C}\in\Delta(54)$, where $\mathrm{C}=\mathrm{S}^2$. 
Hence, the intersection of $T'$ and $\Delta(54)$ is nontrivial and the eclectic flavor group 
$\Omega(1)$ is not given by a semi-direct product of these factors, even though $\Delta(54)$ 
is a normal subgroup of $\Omega(1)$~\cite{Nilles:2020nnc}. 
\item[ii)] Secondly, note that the existence of this discrete $R$-symmetry is linked to a nontrivial automorphy
factor in eq.~\eqref{eq:WgammaTrafo}. Since other nontrivial automorphy factors are possible e.g.\
at specific points in the moduli space of the $T$ modulus, discrete $R$-symmetries are
natural to models with eclectic flavor symmetries. We shall explore in detail this aspect,
associated with the concept of {\it local flavor unification}~\cite{Baur:2019iai}, in a 
forthcoming work~\cite{Nilles:2020tdp}.
\end{enumerate}

On the other hand, as emphasized in ref.~\cite{Chen:2019ewa}, the structure of the K\"ahler 
potential is as important as the superpotential, in particular for flavor phenomenology. The 
K\"ahler potential $K$ is a Hermitian function of the modulus $T$, the chiral superfields $\Phi_n$, 
and their complex conjugates, $\bar{T}$ and $\bar{\Phi}_n$. It must be invariant under the 
traditional flavor symmetry $\Delta(54)$ (since $\int\!\mathrm{d}^2\vartheta\mathrm{d}^2\bar\vartheta$ 
is invariant under $\vartheta \rightarrow \I\,\vartheta$) and transforms covariantly under the 
modular symmetry. The general $\Phi$-independent contribution to the K\"ahler potential is given 
by~\cite{Dixon:1989fj}
\begin{equation}\label{eq:KahlerOfT}
K ~\supset~ -\mathrm{ln}\left(-\I\,T+\I\,\bar{T}\right)\;,
\end{equation}
in Planck units, $M_\mathrm{Pl}=1$. This term is invariant under $\Delta(54)$ and transforms 
under a nontrivial modular transformation $\gamma\in\mathrm{SL}(2,\mathbbm{Z})$ as
\begin{equation}
  -\mathrm{ln}\left(-\I\,T+\I\,\bar{T}\right) ~\stackrel{\gamma}{\longrightarrow}~ -\mathrm{ln}\left(-\I\,T+\I\,\bar{T}\right) + f(T) + \overline{f(T)}\;,
\end{equation}
where $f(T) = \mathrm{ln}(c\,T+d)$. Then, the terms $f(T)+\overline{f(T)}$ are removed by a K\"ahler 
transformation~\cite[ch.23]{Wess:1992cp}, which affects both the K\"ahler potential and the 
superpotential as 
\begin{subequations}\label{eq:Sl2ZPlusKahler}
\begin{eqnarray}
  K           &\stackrel{\gamma}{\longrightarrow}& K + f(T) + \overline{f(T)}
              ~\stackrel{\text{K\"ahler}}{\longrightarrow}~ K\;,\\
  \mathcal W  &\stackrel{\gamma}{\longrightarrow}& (c\,T+d)^{-1}\mathcal{W} 
              \hspace{7mm} ~\stackrel{\text{K\"ahler}}{\longrightarrow}~ (c\,T+d)^{-1} e^{f(T)}\, \mathcal W ~=~ \mathcal W\;,\label{eq:TrafoOfW}
\end{eqnarray}
\end{subequations}
using $f(T) = \mathrm{ln}(c\,T+d)$. This renders the theory modular invariant under 
$\gamma\in\mathrm{SL}(2,\mathbbm{Z})$. Consequently, all additional terms in the K\"ahler potential 
eq.~\eqref{eq:KahlerOfT}, especially those including matter fields, have to be invariant under 
modular transformations. Thus, the transformation properties displayed in eq.~\eqref{eq:TrafoOfW} 
explain why the superpotential $\mathcal W$ has to have modular weight $n=-1$ in 
eq.~\eqref{eq:WgammaTrafo}.

\subsection[Superpotential]{Superpotential}
\label{subsec:Superpotential}

We are interested in building the most general superpotential that is trilinear in the matter 
fields and compatible with all symmetries of the two-dimensional $\mathbbm{T}^2/\Z{3}$ orbifold 
sector: the modular symmetry $\mathrm{SL}(2,\mathbbm{Z})$ and the associated eclectic flavor group 
$\Omega(1)$. In addition, we take into account the standard $\Z{18}^R$ $R$-symmetry related to 
a $\Z{3}$ sublattice rotation in the $\mathbbm{T}^2/\Z{3}$ sector of the full six-dimensional 
orbifold, see ref.~\cite{Nilles:2013lda} and also~\cite{Bizet:2013gf,Nilles:2020tdp}. Using the 
transformation properties of matter fields displayed in table~\ref{tab:Representations}, we find 
that only superpotential terms of the following form are allowed\footnote{Here, we restrict 
ourselves to matter fields from the untwisted and $\theta$ twisted sector. Including
fields from the $\theta^2$ twisted sector leads to
$\mathcal W\, \supset\, \delta^{(0)}(T) \Phi_{\text{\tiny 0}} \Phi_{\nicefrac{-2}{3}} \Phi_{\nicefrac{-1}{3}}
  + \varepsilon^{(0)}(T) \Phi_{\text{\tiny 0}} \Phi_{\nicefrac{-5}{3}} \Phi_{\nicefrac{2}{3}}
  + \zeta^{(2)}(T) \Phi_{\text{\tiny $-1$}} \Phi_{\nicefrac{-5}{3}} \Phi_{\nicefrac{-1}{3}}$,
  where $\delta^{(0)}(T)$ and $\varepsilon^{(0)}(T)$ are modular invariant forms (see eq.~\eqref{eq:ModularInvariantJ}), 
  while $\zeta^{(2)}(T)$ is a modular form with weight $2$ that builds a triplet of $T'$.}
\begin{equation}\label{eq:generalW}
\mathcal{W} ~\supset~ \alpha^{(0)}(T)\, \Phi_{\text{\tiny $-1$}}\,\Phi_{\text{\tiny 0}}\,\Phi_{\text{\tiny 0}} 
                    + \beta^{(1)}(T)\,  \Phi_{\nicefrac{-2}{3}}\, \Phi_{\nicefrac{-2}{3}}\, \Phi_{\nicefrac{-2}{3}}
                    + \gamma^{(4)}(T)\, \Phi_{\nicefrac{-5}{3}}\, \Phi_{\nicefrac{-5}{3}}\, \Phi_{\nicefrac{-5}{3}}\,,
\end{equation}
i.e.\ we find either purely untwisted or purely twisted couplings, where the latter contain only 
matter fields corresponding to twisted strings either without or with oscillator excitations. The 
coupling strengths $\alpha^{(0)}(T)$, $\beta^{(1)}(T)$, and $\gamma^{(4)}(T)$ in 
eq.~\eqref{eq:generalW} are $T$-dependent modular forms due to the modular symmetry 
$\mathrm{SL}(2,\mathbbm{Z})$. Their modular weights have to be $0$, $1$ and $4$, respectively, such 
that the superpotential transforms with modular weight $-1$, as shown in 
eq.~\eqref{eq:Sl2ZPlusKahler}. A modular form $\alpha^{(0)}(T)$ with weight $0$ is modular 
invariant. Thus, $\alpha^{(0)}(T)$ has to be proportional to Klein's $j$ function $j(T)$, which 
is the unique $\mathrm{SL}(2,\mathbbm{Z})$ invariant and holomorphic (away from its cusp) function 
of weight $0$. Hence,
\begin{equation}\label{eq:ModularInvariantJ}
\alpha^{(0)}(T) = \alpha\, j(T)\;,
\end{equation}
where $\alpha\in\mathbbm{C}$ is a free parameter. However, for any value of the K\"ahler modulus 
$T$, the value of $\alpha^{(0)}(T)$ can be chosen freely, from a bottom-up perspective, by 
adjusting the free parameter $\alpha\in\mathbbm{C}$ appropriately. The couplings $\beta^{(1)}(T)$ 
and $\gamma^{(4)}(T)$ have non-vanishing modular weights and, hence, they transform as nontrivial 
$T'$ representations: $\beta^{(1)}(T)$ is a doublet and $\gamma^{(4)}(T)$ is a triplet plus two 
singlets of $T'$. As we will see in section~\ref{sec:Yukawas}, they are fixed uniquely up to an 
overall (complex) factor.

After constructing the relevant couplings $\beta^{(1)}(T)$ and $\gamma^{(4)}(T)$ explicitly in 
section~\ref{sec:Yukawas} using the theory of modular forms, we will build the twisted couplings 
from eq.~\eqref{eq:generalW} step-by-step: First, we only impose the finite modular symmetry $T'$ 
in sections~\ref{subsubsec:WwithT'} and~\ref{subsubsec:WwithT'andOsc}. Afterwards, we impose the 
traditional flavor symmetry $\Delta(54)$ in section~\ref{subsubsec:WwithT'andDelta54}. By doing so, 
we will see that the symmetries of the theory constrain the most general trilinear superpotential 
eq.~\eqref{eq:generalW} such that it is parameterized by only three numbers $c^{(0)}$, $c^{(1)}$ 
and $c^{(4)}\in\mathbbm{C}$. As we shall see, proper field redefinitions allow a further restriction of
these constants to be $c^{(0)},c^{(1)},c^{(4)}\in\mathbbm{R}$. All the rest is fixed by the symmetries 
of the $\mathbbm{T}^2/\Z{3}$ orbifold sector.

\subsubsection[T' properties of modular forms]{\boldmath $T'$ properties of modular forms \unboldmath}
\label{sec:Yukawas}

\begin{table}[t!]
\center
\begin{tabular}{|c||c|c|c|c||c|c|c|c|c|}
\hline
modular              & \multicolumn{8}{c|}{eclectic flavor group $\Omega(1)$}\\
forms                & \multicolumn{4}{c||}{modular $T'$ subgroup} & \multicolumn{4}{c|}{traditional $\Delta(54)$ subgroup}\\
$\hat Y^{(n_Y)}_{\rep{s}}$& \!\!irrep $\rep{s}$\!\! & $\rho_{\rep{s}}(\mathrm{S})$ & $\rho_{\rep{s}}(\mathrm{T})$ & $n_Y$ & \!\!irrep $\rep{r}$\!\! & $\rho_{\rep{r}}(\mathrm{A})$ & $\rho_{\rep{r}}(\mathrm{B})$ & $\rho_{\rep{r}}(\mathrm{C})$ \\
\hline
\hline
$\hat Y^{(1)}_{\rep2''}$ & $\rep2''$ & $\rho_{\rep2''}(\mathrm{S})$ & $\rho_{\rep2''}(\mathrm{T})$ & $1$ & $\rep1$ & $1$               & $1$                   & $1$               \\
\hline
$\hat Y^{(4)}_{\rep1}$   & $\rep1$   & $1$                          & $1$                          & $4$ & $\rep1$ & $1$               & $1$                   & $1$               \\
$\hat Y^{(4)}_{\rep1'}$  & $\rep1'$  & $1$                          & $\omega$                     & $4$ & $\rep1$ & $1$               & $1$                   & $1$               \\
$\hat Y^{(4)}_{\rep3}$   & $\rep3$   & $\rho_{\rep3}(\mathrm{S})$   & $\rho_{\rep3}(\mathrm{T})$   & $4$ & $\rep1$ & $1$               & $1$                   & $1$               \\
\hline
\end{tabular}
\caption{Flavor representations of relevant modular forms $\hat Y^{(n_Y)}_{\rep{s}}(T)$ with 
modular weights $n_Y=1,4$, transforming in the representations $\rep{s}$ of the finite modular 
group $T'$. Here $\omega=\exp(\nicefrac{2\pi\I}{3})$.}
\label{tab:YRepresentations}
\end{table}


Let us denote a general modular form by $\hat Y^{(n_Y)}_{\rep{s}}(T)$ and its modular weight 
by $n_Y \in\mathbbm{N}$. Since we are dealing with the double covering group $T'$ of $A_4$, $n_Y$ 
can be both even or odd~\cite{Liu:2019khw}. First, a modular form is invariant under the 
traditional flavor symmetry, as it only depends on the modulus $T$ of the $\mathbbm{T}^2/\Z{3}$ 
orbifold sector. Second, under a modular transformation $\gamma\in\mathrm{SL}(2,\Z{})$, it 
transforms by definition as a modular form of weight $n_Y$,
\begin{equation}\label{eq:Ymodtrafo}
\hat Y^{(n_Y)}_{\rep{s}}(T) ~\stackrel{\gamma}{\longrightarrow}~ \hat Y^{(n_Y)}_{\rep{s}}\!\left(\tfrac{a\,T+b}{c\,T+d}\right) ~=~ (c\,T+d)^{n_Y} \rho_{\rep{s}}(\gamma) \,\hat Y^{(n_Y)}_{\rep{s}}(T)\,,
\end{equation}
where $\rep{s}$ is the representation of the finite modular group $T'$ under which 
$\hat Y^{(n_Y)}_{\rep{s}}(T)$ transforms.

In addition, it is known that all modular forms with modular weights $n_Y > 1$ can be constructed 
by tensor products of modular forms of weight $n_Y=1$ and the number of independent modular forms 
of a given weight $n_Y$ is finite. Thus, understanding $T'$ modular forms with modular weight $1$ 
provides the information about all possible couplings of the theory.

At weight $1$, there are two independent modular forms of $T'$. A basis is given by~\cite{Liu:2019khw}
\begin{equation}
\hat{e}_1(T) ~:=~ \frac{\eta^3(3\,T)}{\eta(T)} \qquad\mathrm{and}\qquad \hat{e}_2(T) ~:=~ \frac{\eta^3(T/3)}{\eta(T)}\;,
\end{equation}
where $\eta(T)$ is the Dedekind $\eta$-function of the K\"ahler modulus $T$. For later convenience 
we perform the basis change 
\begin{equation}
\label{eq:formsWeight1}
\left(\begin{array}{c} \hat{Y}_1(T) \\ \hat{Y}_2(T) \end{array}\right) ~:=~ 
\left(\begin{array}{cc} -3\sqrt{2} & 0 \\ 3 & 1 \end{array}\right) \left(\begin{array}{c} \hat{e}_1(T) \\ \hat{e}_2(T) \end{array}\right)\;.
\end{equation}
Then, using 
\begin{subequations}
\begin{eqnarray}
\eta\left(T\right) & \stackrel{\mathrm{S}}{\longrightarrow} & \eta\left(-\frac{1}{T}\right) ~=~ \sqrt{-\I\,T}\, \eta(T)\;, \\
\eta\left(T\right) & \stackrel{\mathrm{T}}{\longrightarrow} & \eta\left(T+1\right) ~=~ \exp\left(\frac{\I\,\pi}{12}\right)\, \eta(T)\;,
\end{eqnarray}
\end{subequations}
and
\begin{equation}
\eta^3\left(T+\frac{1}{3}\right) ~=~ \exp\left(\frac{\I\,\pi}{12}\right)\,\eta^3(T) + 3\sqrt{3} \exp\left(-\frac{\I\,\pi}{12}\right)\,\eta^3(9T)\,,
\end{equation}
one can verify that
\begin{subequations}\label{eq:YModularWeight1}
\begin{eqnarray}
\left(\begin{array}{c} \hat{Y}_1(T) \\ \hat{Y}_2(T) \end{array}\right)  & \stackrel{\mathrm{S}}{\longrightarrow} & 
\left(\begin{array}{c} \hat{Y}_1\left(-\frac{1}{T}\right) \\ \hat{Y}_2\left(-\frac{1}{T}\right) \end{array}\right) ~=~ 
(-T)\, \rho_{\rep2''}(\mathrm S) \left(\begin{array}{c} \hat{Y}_1(T) \\ \hat{Y}_2(T) \end{array}\right)\,, \label{eq:ModularSTrafoOfY}\\
\left(\begin{array}{c} \hat{Y}_1(T) \\ \hat{Y}_2(T) \end{array}\right)  & \stackrel{\mathrm{T}}{\longrightarrow} & \left(\begin{array}{c} \hat{Y}_1(T+1) \\ \hat{Y}_2(T+1) \end{array}\right) ~=~ 
\rho_{\rep2''}(\mathrm T)\left(\begin{array}{c} \hat{Y}_1(T) \\ \hat{Y}_2(T) \end{array}\right)\;,\label{eq:ModularTTrafoOfY}
\end{eqnarray}
\end{subequations}
where $(c\,T + d)^{n_Y} = (-1 \cdot T + 0)^1 = (-T)$ is the automorphy factor with weight $n_Y=1$ for the 
modular $\mathrm{S}$ transformation, and 
\begin{equation}
\label{eq:STtrafos2''}
\rho_{\rep2''}(\mathrm S)~:=~-\frac{\I}{\sqrt{3}} \left(\begin{array}{cc} 1 & \sqrt{2} \\ \sqrt{2} & -1 \end{array}\right)\quad\text{and}\quad
\rho_{\rep2''}(\mathrm T)~:=~\left(\begin{array}{cc} \omega & 0 \\ 0 & 1 \end{array}\right) \,.
\end{equation}
Consequently, the couplings $\hat{Y}^{(1)}_{\rep2''}(T):=\left(\hat Y_1(T),\hat Y_2(T)\right)^\mathrm{T}$ 
transform as a doublet $\rep{2}''$ of $T'$, see ref.~\cite{Ishimori:2010au} for notations.

From the structure of the general trilinear superpotential eq.~\eqref{eq:generalW} we know that we 
need the $T'$ modular forms with modular weights $n_Y = 1$ and $n_Y=4$. The later ones correspond 
to the non-vanishing and inequivalent modular forms contained in the tensor product of weight $1$ 
modular forms $\rep2''\otimes\rep2''\otimes\rep2''\otimes\rep2''$. As shown in 
ref.~\cite{Liu:2019khw}, they build the $T'$ representations $\rep1\oplus\rep1'\oplus\rep3$ and are 
given by
\begin{subequations}
\label{eq:formsWeight4}
\begin{eqnarray}
 \hat Y^{(4)}_{\rep1}(T)  & = & 2 \sqrt2 \, \hat Y_1(T)^3 \, \hat Y_2(T) - \hat Y_2(T)^4\,, \label{eq:formsWeight4-1}\\
 \hat Y^{(4)}_{\rep1'}(T) & = & \hat Y_1(T)^4 +  2 \sqrt2 \, \hat Y_1(T) \, \hat Y_2(T)^3\,, \label{eq:formsWeight4-1'}\\
 \hat Y^{(4)}_{\rep3}(T)  & = & \left(\begin{array}{c}
                               \sqrt2\,\hat Y_1(T)^3 \, \hat Y_2(T) + \hat Y_2(T)^4 \\
                               \hat Y_1(T)^4 - \sqrt2\,\hat Y_1(T) \, \hat Y_2(T)^3\\
                               -3 \hat Y_1(T)^2 \, \hat Y_2(T)^2
                             \end{array}\right)\,, \label{eq:formsWeight4-3}
\end{eqnarray}
\end{subequations}
in terms of the basis forms $\hat Y_i(T)$ defined in eq.~\eqref{eq:formsWeight1}. One can readily 
show by using eq.~\eqref{eq:YModularWeight1} that  only $\hat Y^{(4)}_{\rep1}(T)$ and 
$\hat Y^{(4)}_{\rep1'}(T)$ acquire the automorphy factor $(-T)^4$ under the modular 
$\mathrm{S}$ transformation, while $\hat Y^{(4)}_{\rep1}(T)$ is left invariant by $\mathrm T$ and 
$\hat Y^{(4)}_{\rep1'}(T)$ gets the phase $\omega$. This implies, according to 
eq.~\eqref{eq:Ymodtrafo}, that $\hat Y^{(4)}_{\rep1}(T)$ and $\hat Y^{(4)}_{\rep1'}(T)$ build the 
$\rep1$ and $\rep1'$ representations of $T'$, respectively. Finally, the triplet 
$\hat Y^{(4)}_{\rep3}(T)$ transforms under $\mathrm{S}$ and $\mathrm{T}$ according to 
eq.~\eqref{eq:Ymodtrafo} with
\begin{equation}
  \rho_{\rep3}(\mathrm S)~=~ \frac13 \left( \begin{array}{ccc}-1 &  2 & -2 \\
                                                               2 & -1 & -2 \\
                                                              -2 & -2 & -1 
                                            \end{array}\right)\,,\qquad
  \rho_{\rep3}(\mathrm T)~=~ \left( \begin{array}{ccc} 1 & 0      & 0 \\
                                                       0 & \omega & 0 \\
                                                       0 & 0      & \omega^2 
                                            \end{array}\right)\,.
\end{equation}
Consequently, $\hat Y^{(4)}_{\rep3}(T)$ builds a representation $\rep3$ of $T'$. The $T'$ (and 
$\Delta(54)$) representations of all relevant modular forms are summarized in 
table~\ref{tab:YRepresentations}.

\subsubsection[T' modular invariant superpotential for matter fields with n=-2/3]{\boldmath $T'$ modular invariant superpotential for matter fields with $n=\nicefrac{-2}{3}$ \unboldmath}
\label{subsubsec:WwithT'}

Let us construct now the most general trilinear superpotential of three copies of twisted matter 
fields $\Phi_{\nicefrac{-2}{3}}^i = (X_i, Y_i, Z_i)^\mathrm{T}$, $i=1,2,3$. These fields correspond 
to $\theta$-twisted strings without oscillator excitations. In this case, the modular weight $n_Y$ 
of the coupling strength $\hat Y^{(n_Y)}_{\rep{s}}(T)$ and the modular weights $n=\nicefrac{-2}{3}$ 
of the three twisted matter fields $\Phi_{\nicefrac{-2}{3}}^i$, $i=1,2,3$, have to fulfill the 
condition $n_Y + 3 \cdot (\nicefrac{-2}{3})=-1$, see eq.~\eqref{eq:WgammaTrafo}. Thus, we need 
$n_Y=1$ and the coupling strength is given by the $T'$ doublet $\hat Y_{\rep2''}^{(1)}(T)$ in 
eq.~\eqref{eq:formsWeight1}. Then, a trilinear coupling of twisted matter fields 
$\Phi_{\nicefrac{-2}{3}}^i$ originates from the trivial singlet $\rep{1}$ resulting from the tensor 
products of $T'$ representations
\begin{equation}\label{eq:TensorProductThetaSectorNoOsci}
\rep{1} ~\subset~ \rep{2}'' \otimes \left( \rep{2}' \oplus \rep{1} \right) \otimes \left( \rep{2}' \oplus \rep{1} \right) \otimes \left( \rep{2}' \oplus \rep{1} \right)\;,
\end{equation}
corresponding to 
\begin{equation}
\mathcal{W} ~\supset~ \left(\begin{array}{c} \hat{Y}_1(T) \\ \hat{Y}_2(T) \end{array}\right) \otimes \left(\begin{array}{c} X_1 \\ Y_1 \\ Z_1 \end{array}\right) \otimes \left(\begin{array}{c} X_2 \\ Y_2 \\ Z_2 \end{array}\right) \otimes \left(\begin{array}{c} X_3 \\ Y_3 \\ Z_3 \end{array}\right)\;,
\end{equation}
see table~\ref{tab:Representations}, and we assume that only the product of the three different 
twisted triplets, $\Phi_{\nicefrac{-2}{3}}^1\,\Phi_{\nicefrac{-2}{3}}^2\,\Phi_{\nicefrac{-2}{3}}^3$, 
is allowed, for example, by gauge invariance. Then, writing out the tensor 
products~(\ref{eq:TensorProductThetaSectorNoOsci}) explicitly using ref.~\cite{Ishimori:2010au} 
(with $p=\I$, $p_1=1$ and $p_2=-1$), we obtain four independent $T'$ singlets 
$\mathcal{W}_a(T, X_i, Y_i, Z_i)$, given by eq.~\eqref{eq:WtermsNoOscillators} in 
appendix~\ref{sec:Wterms}. Therefore, at first sight, the trilinear superpotential 
$\mathcal{W}(T,X_i,Y_i,Z_i)$ of the K\"ahler modulus $T$ and the twisted fields 
$(X_i,Y_i,Z_i)^\mathrm{T}$ contains four independent coefficients $c_a\in \mathbbm{C}$, 
$a=1,\ldots,4$ (or modular invariant functions $c_a(T)$, cf.\ the discussion around 
eq.~\eqref{eq:ModularInvariantJ}),
\begin{equation}\label{eq:TprimeSuperpotential}
\mathcal{W} ~\supset~ \sum_{a=1}^4 c_a\,\mathcal{W}_a(T, X_i, Y_i, Z_i)\;.
\end{equation}
In other words, the superpotential eq.~\eqref{eq:TprimeSuperpotential} is the most general 
trilinear superpotential of twisted fields with modular weights $n=\nicefrac{-2}{3}$ if one assumes 
invariance only under the modular symmetry $T'$. It is parameterized by four (modular invariant) 
coefficients $c_a$. As we shall see in section~\ref{subsubsec:WwithT'andDelta54}, these four
coefficients are reduced to one, after imposing invariance under the traditional flavor symmetry 
$\Delta(54)$.

\subsubsection[T' modular invariant superpotential for matter fields with n=-5/3]{\boldmath $T'$ modular invariant superpotential for matter fields with $n=\nicefrac{-5}{3}$ \unboldmath}
\label{subsubsec:WwithT'andOsc}

Next, we construct the most general trilinear superpotential of three copies of twisted matter 
fields $\Phi_{\nicefrac{-5}{3}}^i = (\tilde{X}_i, \tilde{Y}_i, \tilde{Z}_i)^\mathrm{T}$, $i=1,2,3$, 
again under the assumption that only the product 
$\Phi_{\nicefrac{-5}{3}}^1\,\Phi_{\nicefrac{-5}{3}}^2\,\Phi_{\nicefrac{-5}{3}}^3$ is allowed by 
gauge invariance. From a string point of view, these fields originate from $\theta$-twisted strings 
with oscillator excitations. As anticipated, the couplings are given in this case by modular forms 
of weight $n_Y=4$ such that $n_Y + 3 \cdot (\nicefrac{-5}{3})=-1$ is the modular weight of the 
superpotential. 

The three triplets of twisted matter fields $\Phi_{\nicefrac{-5}{3}}^i$ transform in the $T'$ 
representations $\rep2'\oplus\rep1$, see table~\ref{tab:Representations}. Thus, $T'$ invariant 
couplings must result from the $T'$ tensor products
\begin{equation}
\label{eq:tensorproducts-withoscillators}
\rep{1} ~\subset~ \left( \rep1\oplus\rep1'\oplus\rep3 \right) \otimes \left( \rep{2}' \oplus \rep{1} \right) \otimes 
                  \left( \rep{2}' \oplus \rep{1} \right) \otimes \left( \rep{2}' \oplus \rep{1} \right)\;,
\end{equation}
corresponding to 
\begin{equation}\label{eq:GeneralFormSuperpotentialWithOsci}
\mathcal{W} ~\supset~ \left( \hat Y^{(4)}_{\rep1}(T) \oplus \hat Y^{(4)}_{\rep1'}(T) \oplus \hat Y^{(4)}_{\rep3}(T) \right) \otimes 
                      \left(\begin{array}{c} \tilde X_1 \\ \tilde Y_1 \\ \tilde Z_1 \end{array}\right) \otimes 
                      \left(\begin{array}{c} \tilde X_2 \\ \tilde Y_2 \\ \tilde Z_2 \end{array}\right) \otimes 
                      \left(\begin{array}{c} \tilde X_3 \\ \tilde Y_3 \\ \tilde Z_3 \end{array}\right)\;.
\end{equation}
Here, the modular forms $\hat Y^{(4)}_{\rep s}(T)$ of weight $n_Y=4$ are given in 
eq.~\eqref{eq:formsWeight4}. These tensor products yield seven independent $T'$ invariant couplings 
$\widetilde{\mathcal W}_a(T, \tilde X_i,\tilde Y_i,\tilde Z_i)$, $a=1,\ldots,7$, given in
eq.~\eqref{eq:WtermsOscillators} of appendix~\ref{sec:Wterms}. Then, the trilinear superpotential 
of three copies of twisted matter fields $\Phi_{\nicefrac{-5}{3}}^i$, $i=1,2,3$, reads
\begin{equation}\label{eq:TwistedOsciGeneralForm}
\mathcal W \supset \sum_{a=1}^7 \tilde{c}_a ~ \widetilde{\mathcal W}_a(T, \tilde X_i,\tilde Y_i,\tilde Z_i)\,,
\end{equation}
where $\tilde{c}_a$, $a=1,~\ldots,7$, denote seven independent coefficients (i.e.\ modular 
invariant functions as discussed around eq.~\eqref{eq:ModularInvariantJ}). We shall show shortly 
that the traditional flavor symmetry $\Delta(54)$ invariance further constrains  these superpotential 
couplings, reducing the number of free coefficients to single one.

\subsubsection[Restrictions from Delta(54)]{\boldmath Restrictions from $\Delta(54)$ \unboldmath}
\label{subsubsec:WwithT'andDelta54}

Since $T'$ represents only the modular subgroup of the full eclectic flavor group $\Omega(1)$ of 
the $\mathbbm{T}^2/\Z{3}$ orbifold sector, we must impose additional constraints to arrive at a 
consistent superpotential. These constraints arise from the $\Delta(54)$ traditional flavor group. 
As shown in table~\ref{tab:Representations}, $\mathcal W$ must transform as a nontrivial singlet 
$\rep1'$ of $\Delta(54)$. While the untwisted trilinear couplings in eq.~\ref{eq:generalW} satisfy 
this condition automatically, one must identify the linear combinations of the twisted couplings, 
i.e.\ $\mathcal{W}_a(T, X_i, Y_i, Z_i)$ in eq.~\eqref{eq:TprimeSuperpotential} and 
$\widetilde{\mathcal W}_a(T, \tilde X_i,\tilde Y_i,\tilde Z_i)$ in eq.~\eqref{eq:TwistedOsciGeneralForm}, 
that are invariant under the $\Delta(54)$ generators $\mathrm{A}$ and $\mathrm{B}$ and transform 
covariantly under the $R$-symmetry generator $\mathrm{C}$.

We find that consistency with $\Delta(54)$ restricts the coefficients $c_a$ in
eq.~\eqref{eq:TprimeSuperpotential} to be equal, reducing these terms in the superpotential to
\begin{subequations}
\label{eq:Z3superpotential}
\begin{eqnarray}
\mathcal W(T,X_i,Y_i,Z_i) & \supset & c^{(1)}\, \Big[\hat{Y}_2(T) \big( X_1\,X_2\,X_3 + Y_1\,Y_2\,Y_3 + Z_1\,Z_2\,Z_3\big)  \label{eq:unsuppressedtwisted}\\
                          &         & \hspace{8mm} - \frac{\hat{Y}_1(T)}{\sqrt{2}} \big( X_1\,Y_2\,Z_3 + X_1\,Y_3\,Z_2 + X_2\,Y_1\,Z_3 \label{eq:suppressedtwisted}\\
                          &         & \hspace{22mm} +\, X_3\,Y_1\,Z_2 + X_2\,Y_3\,Z_1 + X_3\,Y_2\,Z_1\big)\Big]\;, \nonumber
\end{eqnarray}
\end{subequations}
where $c^{(1)}=c_a$ for $a=1,\ldots,4$ can be chosen to be a constant. Interestingly, the relative 
coupling strength $-\sqrt{2}\hat{Y}_2(T)/\hat{Y}_1(T)$ of twisted matter fields localized at the 
same orbifold fixed point (e.g.\ $X_1\,X_2\,X_3$) and twisted matter fields localized at three 
different orbifold fixed points (e.g.\ $X_1\,Y_2\,Z_3$) is completely fixed by the eclectic flavor 
symmetry $\Omega(1)$ without any free parameter. Moreover, note that one can absorb the phase of 
the overall constant $c^{(1)}$ in eq.~\eqref{eq:Z3superpotential} into a redefinition of the fields 
$X_i$, $Y_i$, $Z_i$, such that we can set $c^{(1)} \in \mathbbm{R}$. 

Similarly, we find that $\Delta(54)$ covariance of eq.~\eqref{eq:TwistedOsciGeneralForm} requires
$c^{(4)}=\tilde c_1=-\tilde c_2=c_3$ and $\tilde c_{4,5,6,7}=0$, which leads to the 
superpotential contribution
\begin{eqnarray}
\label{eq:Z3superpotential-withoscillators}
\mathcal W(T,\tilde X_i,\tilde Y_i,\tilde Z_i) & \supset & c^{(4)}\, \hat Y^{(4)}_{\rep1'}(T) 
               \Big(  \tilde X_1\,\tilde Y_3\,\tilde Z_2 - \tilde X_1\,\tilde Y_2\,\tilde Z_3 + \tilde X_2\,\tilde Y_1\,\tilde Z_3\\ 
 & & \hspace{2cm} -\tilde X_2\,\tilde Y_3\,\tilde Z_1 + \tilde X_3\,\tilde Y_2\,\tilde Z_1 - \tilde X_3\,\tilde Y_1\,\tilde Z_2\Big)\;. \nonumber
\end{eqnarray}
Similar to eq.~\eqref{eq:Z3superpotential}, the complex phase of the overall constant $c^{(4)}$ can 
be absorbed by a field redefinition such that $c^{(4)} \in \mathbbm{R}$. Note that 
eq.~\eqref{eq:Z3superpotential-withoscillators} is antisymmetric in the exchange of 
$\Phi_{\nicefrac{-5}{3}}^i = (\tilde{X}_i, \tilde{Y}_i, \tilde{Z}_i)^\mathrm{T}$ and 
$\Phi_{\nicefrac{-5}{3}}^j = (\tilde{X}_j, \tilde{Y}_j, \tilde{Z}_j)^\mathrm{T}$, for $i,j=1,2,3$ 
and $i\neq j$. Furthermore, the coupling strength $\hat Y^{(4)}_{\rep1'}(T)$ of this interaction is 
given by eq.~\eqref{eq:formsWeight4-1'}. 

A couple of remarks on the twisted superpotential are in order. First, we recall that 
$\mathrm{Im}(T)$ corresponds to the volume of the $\mathbbm{T}^2/\Z{3}$ orbifold sector. Then, in 
the so-called large-volume limit defined by $T\rightarrow \I\infty$, the superpotential couplings 
become
\begin{equation}
\hat{Y}_1(T) ~\rightarrow~ 0  \qquad\mathrm{and}\qquad \hat{Y}_2(T) ~\rightarrow~ 1\;.
\end{equation}
Hence, this yields $\hat Y^{(4)}_{\rep1'}(T) \rightarrow 0$. We note that this limit reproduces 
the intuitive result that couplings of twisted strings are suppressed if the strings have to 
stretch in order to meet in the compactified dimensions and then join together: the couplings in 
eq.~\eqref{eq:unsuppressedtwisted} of three twisted strings localized at the same fixed point of 
the $\mathbbm{T}^2/\Z{3}$ orbifold sector are unsuppressed (e.g.\ for $X_1\,X_2\,X_3$), while the 
couplings in eqs.~\eqref{eq:suppressedtwisted} and~\eqref{eq:Z3superpotential-withoscillators} of 
three twisted strings localized at three different fixed points vanish (e.g.\ for $X_1\,Y_2\,Z_3$).

Secondly, we realize that trilinear interactions of twisted matter fields 
$\Phi_{\nicefrac{-5}{3}}^i= (\tilde{X}_i, \tilde{Y}_i, \tilde{Z}_i)^\mathrm{T}$ are excluded in 
eq.~\eqref{eq:Z3superpotential-withoscillators} if the three twisted matter fields are localized at 
the same orbifold fixed point: In contrast to the interactions in eq.~\eqref{eq:unsuppressedtwisted}, 
there are no terms analogous to, for example, $X_1\,X_2\,X_3$. At first sight, this might seem to 
contradict the intuitive picture of string interactions on orbifolds. However, it is known in 
string theory~\cite{Font:1988tp} that twisted strings localized at the same \Z3 orbifold fixed 
point must satisfy the condition that in each coupling the number of holomorphic oscillator 
excitations must equal the number of anti-holomorphic excitations modulo six. This string 
constraint is known as ``rule 4'', see refs.~\cite{Font:1988tp,Kobayashi:2011cw}. In our case, 
each twisted string carries one holomorphic oscillator excitation and there are no anti-holomorphic 
excitations. Thus, a coupling like $\tilde{X}_1\,\tilde{X}_2\,\tilde{X}_3$ is forbidden by rule 4. 
Interestingly, our superpotential eq.~\eqref{eq:Z3superpotential-withoscillators} shows that rule 4 
is automatically satisfied if the theory is $\Omega(1)$ invariant.

\subsection[K\"ahler potential]{\boldmath K\"ahler potential\unboldmath}
\label{subsec:Kahlerpotential}

It is known that the leading order K\"ahler potential of general matter fields $\Phi_n$ with 
modular weights $n$ originating from string compactifications on Abelian orbifolds has the 
form~\cite{Dixon:1989fj}
\begin{equation}
\label{eq:MatterKaehlerFirstAnsatz}
  K ~\supset~ \sum_{\Phi_n}\, (-\I T + \I \bar T)^{n} |\Phi_{n}|^2\,.
\end{equation}
Here, additional (gauge) charges are assumed that forbid terms like 
$\Phi_{n,1} \bar\Phi_{n,2}+\Phi_{n,2} \bar\Phi_{n,1}$ combining different matter fields $\Phi_{n,1}$ 
and $\Phi_{n,2}$. As suggested in ref.~\cite{Chen:2019ewa}, invariance under the modular group 
alone does not fix the structure of eq.~\eqref{eq:MatterKaehlerFirstAnsatz}. From a bottom-up 
perspective, the K\"ahler potential can in principle receive unsuppressed contributions from 
modular forms $\hat Y^{(n_Y)}_{\rep s}(T)$. These extra terms can significantly alter the 
phenomenological predictions that have been obtained by using just the standard K\"ahler potential 
eq.~\eqref{eq:MatterKaehlerFirstAnsatz}. To be specific, such terms can introduce nontrivial 
mixtures in the quark and lepton sectors.

Based on these observations, we follow ref.~\cite{Chen:2019ewa} and generalize 
eq.~(\ref{eq:MatterKaehlerFirstAnsatz}) to the following ansatz for the K\"ahler potential of matter 
fields:
\begin{equation}
\label{eq:MatterKaehlerAnsatz}
  K ~\supset~ \sum_{\Phi_n} \sum_{n_Y\geq0} (-\I T + \I \bar T)^{n+n_Y} 
         \sum_{a} \kappa_a^{(n_Y)} \left[\hat Y^{(n_Y)}_{\rep s}(T) \otimes \Phi_{n} \otimes \left(\hat Y^{(n_Y)}_{\rep s}(T)\right)^* \otimes\bar\Phi_{n}\right]_{\rep1,a}\,,
\end{equation}
where we sum over all fields $\Phi_n$ with modular weights $n=0,-1,\nicefrac{-2}{3},\nicefrac{-5}{3}$ 
from the $\mathbbm{T}^2/\Z{3}$ orbifold sector and we introduce coefficients 
$\kappa_a^{(n_Y)}\in\mathbbm{R}$. Moreover, we sum over all modular weights $n_Y\in\mathbbm{N}$ of 
the modular forms $\hat Y^{(n_Y)}_{\rep s}(T)$ and all ($\Delta(54)$ and $T'$) singlet contractions, 
labeled by the index $a$. Here, we also allow for $n_Y=0$, taking $\hat Y^{(0)}_{\rep s}=1$ 
in this case.\footnote{Formally $\hat Y^{(0)}_{\rep s}\propto j(T)$, however, following
our discussion around eq.~\eqref{eq:ModularInvariantJ}, it is possible to fix $\hat Y^{(0)}_{\rep s}=1$.}
Furthermore, for each $n_Y$ we consider implicitly all admissible $T'$ representations $\rep s$ 
of $\hat Y^{(n_Y)}_{\rep s}$. Since untwisted matter fields are $\Delta(54)$ and $T'$ singlets, the 
structure of their K\"ahler potential is rather trivial and we can skip their discussion in 
the following.

By construction and considering that $\left[\ldots\right]_{\rep1,a}$ refers to singlet contractions, 
the ansatz~\eqref{eq:MatterKaehlerAnsatz} for the matter K\"ahler potential is $\Delta(54)$ and $T'$ 
invariant. Moreover, according to our discussion in section~\ref{sec:action} the matter K\"ahler 
potential must be invariant under modular transformations $\mathrm{SL}(2,\Z{})$ as well. In detail,
under an arbitrary modular transformation $\gamma\in\mathrm{SL}(2,\mathbbm Z)$, we see that the 
first factor in eq.~\eqref{eq:MatterKaehlerAnsatz} transforms as
\begin{equation}
\label{eq:trafoOfKaehlerAnsatzprefactor}
(-\I T + \I \bar T)^{n+n_Y} \stackrel{\gamma}{\longrightarrow} (c\,T + d)^{-n-n_Y} (c\,\bar T + d)^{-n-n_Y} (-\I T + \I \bar T)^{n+n_Y}\,.
\end{equation}
According to eqs.~\eqref{eq:ModularTransformationOfPhi} and~\eqref{eq:Ymodtrafo}, the $T'$ singlet 
contractions $\left[\ldots\right]_{\rep1,a}$ in eq.~\eqref{eq:MatterKaehlerAnsatz} transform 
precisely with the correct automorphy factors to compensate the factors in 
eq.~\eqref{eq:trafoOfKaehlerAnsatzprefactor}. Hence, the K\"ahler potential 
eq.~\eqref{eq:MatterKaehlerAnsatz} is invariant under both, $\mathrm{SL}(2,\Z{})$ and the finite 
modular group $T'$. We point out that invariance under only $T'$ and $\Delta(54)$ would allow 
additional terms involving modular forms of different modular weights. However, these terms 
are forbidden by the automorphy factors of $\mathrm{SL}(2,\Z{})$.

Let us now explore more explicitly the K\"ahler potential of a twisted matter field that follows 
from the ansatz~\eqref{eq:MatterKaehlerAnsatz}.
For a twisted matter field $\Phi_n$, the K\"ahler potential is independent of the specific modular 
weight $n$. Thus, we can choose for example a triplet of $\theta$-twisted matter fields 
$\Phi_{\nicefrac{-2}{3}}=(X,Y,Z)^\mathrm{T}$ with $n=\nicefrac{-2}{3}$. In this case, just 
demanding that $K$ be Hermitian restricts the matter K\"ahler potential to the general form 
\begin{subequations}
\label{eq:TwisttedKaehlerInvariants}
\begin{eqnarray}
\label{eq:TwisttedKaehlerInvariantsFull}
  K &\!\!\supset\!&\!\! \sum_{n_Y\geq0} (-\I T + \I \bar T)^{n+n_Y} \bigg[ A^{(n_Y)}_1(T,\bar T) |X|^2 + A^{(n_Y)}_2(T,\bar T) |Y|^2 + A^{(n_Y)}_3(T,\bar T) |Z|^2 \quad\\
\label{eq:TwisttedKaehlerInvariantsT'1}
    &             &\hspace{3.2cm}+\, A^{(n_Y)}_4(T,\bar T)\left(X\bar Y + \bar X Y\right) + A^{(n_Y)}_5(T,\bar T)\left(X\bar Z + \bar X Z\right) \\
\label{eq:TwisttedKaehlerInvariantsT'2}
    &             &\hspace{3.2cm}+\, A^{(n_Y)}_6(T,\bar T)\left(Y\bar Z + \bar Y Z\right)\bigg]\,, 
\end{eqnarray}
\end{subequations}
where, compared to eq.~\eqref{eq:MatterKaehlerAnsatz}, the real functions $A^{(n_Y)}_m(T,\bar T)$, 
$m=1,\ldots,6$, depend on $\kappa_a^{(n_Y)}$, the modular forms $\hat Y^{(n_Y)}_{\rep s}(T)$ and 
the Clebsch--Gordan coefficients of the tensor products. This parameterization of $K$ is beneficial 
in order to see the non-diagonal terms $A^{(n_Y)}_m(T,\bar T)$ for $m=4,5,6$ in 
eqs.~\eqref{eq:TwisttedKaehlerInvariantsT'1} and~\eqref{eq:TwisttedKaehlerInvariantsT'2}. From a 
phenomenological point of view, independently of the form of the superpotential, these non-diagonal 
terms can lead to mixed mass eigenstates and, hence, nontrivial textures in the mixing matrices 
for $\Phi_n$ corresponding to quark or lepton fields. However, the functions $A_m^{(n_Y)}$ are 
constrained by imposing invariance under all symmetries of the theory, as we discuss next. We 
proceed in two steps: first, we only impose modular invariance under $\mathrm{SL}(2,\mathbbm Z)$ 
and $T'$ and, in a second step, we consider restrictions from the traditional flavor symmetry 
$\Delta(54)$. By doing so, we will uncover some of the advantages of the eclectic approach to 
flavor symmetries.

\subsubsection[T' invariant K\"ahler potential]{\boldmath $T'$ invariant K\"ahler potential \unboldmath}
\label{subsubsec:KwithT'}

Let us consider first only $T'$ invariance and compute explicitly the resulting K\"ahler potential 
of a twisted matter field $\Phi_n=(X,Y,Z)^\mathrm{T}$ for some specific modular forms 
$\hat Y^{(n_Y)}_{\rep s}$ of modular weights $n_Y$.

For $n_Y=0$ (i.e.\ in the absence of modular forms $\hat Y^{(n_Y)}_{\rep s}$), we find that the 
general ansatz~\eqref{eq:MatterKaehlerAnsatz} for the K\"ahler potential of twisted matter fields 
$\Phi_n$ is given by
\begin{subequations}\label{eq:Kaehlerw1}
\begin{eqnarray}
K ~\supset~ \left(-\I\,T+\I\,\bar{T}\right)^{n}\,\bigg[\kappa^{(0)}_1 |X|^2  
             &+& \frac{1}{2}\left(\kappa^{(0)}_1+\kappa^{(0)}_2\right)\,\left(|Y|^2 + |Z|^2\right)\\
             &+& \left. \frac{1}{2}\left(\kappa^{(0)}_1-\kappa^{(0)}_2\right)\,\left(Y \bar{Z} + \bar{Y} Z\right)\right]\;.\label{eq:Kaehlerw1line2}
\end{eqnarray}
\end{subequations}
These terms originate from the $T'$ tensor product $(\rep{2}'\oplus\rep{1})\otimes(\rep{2}''\oplus\rep{1})$ 
that yields two independent invariants with coefficients $\kappa^{(0)}_1$ and $\kappa^{(0)}_2$. 
Comparing with eq.~\eqref{eq:TwisttedKaehlerInvariants}, we realize that here, $A_1^{(0)}$, $A_2^{(0)}$, 
$A_3^{(0)}$ and $A_6^{(0)}$ are non-vanishing constants. That is, considering only 
$T'$ invariance, there is a non-diagonal mixing among fields in this case, see 
eq.~\eqref{eq:Kaehlerw1line2}. As we shall see shortly, imposing in addition the $\Delta(54)$ 
traditional flavor symmetry eliminates this mixing.

For $n_Y=1$, the general ansatz~\eqref{eq:MatterKaehlerAnsatz} depends on the modular forms
$\hat Y^{(1)}_{\rep 2''}(T)$ defined in eq.~\eqref{eq:formsWeight1}. Considering the three $T'$ 
invariants contained in the tensor product of $|\hat Y^{(1)}_{\rep 2''}(T)\otimes\Phi_n|^2$ 
(related to $|\rep2''\otimes(\rep{2}'\oplus\rep{1})|^2=|\rep1\oplus\rep3\oplus\rep2''|^2$), we find
\begin{eqnarray}
\label{eq:Kaehlerw2}
K &\supset& \left(-\I\,T+\I\,\bar{T}\right)^{n+1}\,\frac12
                            \bigg[ \left((\kappa^{(1)}_1+\kappa^{(1)}_2)\,|\hat Y_1(T)|^2 + 2\,\kappa^{(1)}_2|\hat Y_2(T)|^2 \right) |X|^2  \\
          &&                \hspace{28mm}   + \,\frac12\left((\kappa^{(1)}_1 + \kappa^{(1)}_2)\,|\hat Y_2(T)|^2 + 2\,\kappa^{(1)}_2|\hat Y_1(T)|^2 \right) |Y + Z|^2 \nonumber\\
          &&                \hspace{28mm}   + \,\frac12\kappa^{(1)}_3 \left(|\hat Y_1(T)|^2+|\hat Y_2(T)|^2 \right) |Y - Z|^2 \nonumber\\
          &&                \hspace{28mm}   + \,\sqrt2 \,\text{Re}\left\{\hat Y_1^*(T) \hat Y_2(T) (\kappa^{(1)}_1-\kappa^{(1)}_2)\bar{X}(Y+Z)\right\} 
                            \bigg]\;. \nonumber
\end{eqnarray}
Comparing this K\"ahler potential with the general scheme eq.~\eqref{eq:TwisttedKaehlerInvariants},
we find that, if only the $T'$ modular flavor symmetry is taken into account, admitting modular 
forms with the lowest modular weight in the K\"ahler potential leads to non-vanishing $A_m^{(1)}$ 
for all $m=1,\ldots,6$, which in turn yield nontrivial mixings. Furthermore, the explicit expressions 
of the functions $A_m^{(1)}$ do not seem to have a simple connection to  the constants $A_m^{(0)}$ 
of eq.~\eqref{eq:Kaehlerw1}. 
These findings reveal that the $T'$ finite modular symmetry is not very restrictive for the 
K\"ahler potential. In general, all coefficients $A^{(n_Y)}_m(T,\bar T)$ in 
eq.~\eqref{eq:TwisttedKaehlerInvariants} appear at some modular weight $n_Y$, resulting in all 
the possible non-diagonal mixings.

\subsubsection[Restrictions from Delta(54)]{\boldmath Restrictions from $\Delta(54)$ \unboldmath}
\label{subsubsec:KwithT'andDelta54}

The traditional flavor symmetry $\Delta(54)$ includes the $\Z{3}^{\mathrm{(PG)}}\!\times\Z{3}$ point 
group and space group symmetries, see eq.~\eqref{eq:pointgroupselectionrule}. Thus, demanding 
invariance first under $\Z{3}^{\mathrm{(PG)}}\!\times\Z{3}$ implies that the K\"ahler potential 
eq.~\eqref{eq:TwisttedKaehlerInvariants} reduces to the terms contained in 
eq.~\eqref{eq:TwisttedKaehlerInvariantsFull}, i.e.\ it has to be a function of $|X|^2$, $|Y|^2$ and 
$|Z|^2$ only: $A^{(n_Y)}_m=0$ for $m=4,5,6$. In addition, applying the $\Delta(54)$ transformation 
$\rho(\mathrm{A})$ from eq.~\eqref{eq:Delta54Generators} on the triplet $\Phi_n$ interchanges the 
twisted matter fields $X$, $Y$ and $Z$. Thus, the terms in eq.~\eqref{eq:TwisttedKaehlerInvariants} 
are further constrained to 
\begin{equation}\label{eq:FinalMatterKaehlerAnsatz}
K ~\supset~ \sum_{n_Y\geq0} (-\I T + \I \bar T)^{n+n_Y} A^{(n_Y)}_1(T,\bar T)\,\left( |X|^2 + |Y|^2 + |Z|^2\right)\,,
\end{equation}
where $A_1^{(n_Y)} = A_2^{(n_Y)} = A_3^{(n_Y)}$. Hence, we observe that, in contrast to the 
(finite) modular symmetry only, the traditional flavor symmetry $\Delta(54)$ forbids all 
non-diagonal terms.

Notice that $|\Phi_n|^2 = |X|^2 + |Y|^2 + |Z|^2$ is the unique $\Delta(54)$ and $T'$ singlet from 
$\bar{\Phi}_n \otimes \Phi_n$. On the other hand, under a general modular transformation 
$\gamma\in\mathrm{SL}(2,\mathbbm Z)$, $|\Phi_n|^2$ transforms with an automorphy factor,
\begin{equation}
|X|^2 + |Y|^2 + |Z|^2 \stackrel{\gamma}{\longrightarrow} \big|(c\,T + d)^{n}\big|^2\,\left( |X|^2 + |Y|^2 + |Z|^2\right)\;,
\end{equation}
using $\rho_{\rep{3}_n}\!(\mathrm{\gamma})^\dagger \rho_{\rep{3}_n}\!(\mathrm{\gamma})=\Id$ which 
follows from $\rho_{\rep{3}_n}\!(\mathrm{S})^\dagger \rho_{\rep{3}_n}\!(\mathrm{S})=\rho_{\rep{3}_n}\!(\mathrm{T})^\dagger \rho_{\rep{3}_n}\!(\mathrm{T})=\Id$.
Consequently, $A^{(n_Y)}_1(T,\bar T)$ is restricted to be a trivial singlet $\rep1$ of $T'$ 
transforming under $\gamma\in\mathrm{SL}(2,\Z{})$ as
\begin{equation}
A^{(n_Y)}_1(T,\bar T) \stackrel{\gamma}{\longrightarrow} \big|(c\,T+d)^{n_Y}\big|^2 A^{(n_Y)}_1(T,\bar T)\;.
\end{equation}
Then, the K\"ahler contributions eq.~\eqref{eq:FinalMatterKaehlerAnsatz} are modular invariant 
after taking into account eq.~\eqref{eq:trafoOfKaehlerAnsatzprefactor}. Hence, comparing 
eq.~\eqref{eq:FinalMatterKaehlerAnsatz} with our original ansatz eq.~\eqref{eq:MatterKaehlerAnsatz}, 
we find that
\begin{equation}
 A^{(n_Y)}_1(T,\bar T) ~=~ \sum_a \kappa_a^{(n_Y)}\left|\hat Y^{(n_Y)}_{\rep s}(T)\right|^2_{\rep1,a}\,.
\end{equation}

In summary, we can conclude that the most general K\"ahler potential bilinear in twisted matter 
fields, compatible with the eclectic flavor group $\Omega(1)$, is given by
\begin{subequations}
\begin{eqnarray}
K &\supset& \sum_{\Phi_n} \left(\sum_{n_Y\geq0} \left(-\I\,T+\I\,\bar{T}\right)^{n+n_Y} 
                        \sum_a \kappa_a^{(n_Y)}\left|\hat Y^{(n_Y)}_{\rep s}(T)\right|^2_{\rep1,a} \right) \left|\Phi_n\right|^2\\
\label{eq:DiagonalKaehler}
                    &=:& \sum_{\Phi_n} g_n(T,\bar T)\,\left|\Phi_n\right|^2\,, 
\end{eqnarray}
\end{subequations}
where $g_n(T,\bar T)$ is defined as the element of the diagonal K\"ahler metric corresponding to 
the matter field $\Phi_n$. From its definition, one can explicitly compute $g_n(T,\bar T)$ for each 
matter field evaluating the modular forms with different modular weights $n_Y$. For example, for 
$n_Y=0,1,2$ we obtain
\begin{subequations}\label{eq:ginK}
\begin{eqnarray}
  g_n(T, \bar{T})\!\!\!&=&\!\!\! \kappa_1^{(0)} \left(-\I\,T+\I\,\bar{T}\right)^{n} \label{eq:ginK0}\\
                       &+&\!\!\! \kappa_1^{(1)} \left(-\I\,T+\I\,\bar{T}\right)^{n+1} \left(|\hat Y_1(T)|^2 + |\hat Y_2(T)|^2\right) \label{eq:ginK1}\\ 
                       &+&\!\!\! \kappa_1^{(2)} \left(-\I\,T+\I\,\bar{T}\right)^{n+2} \left(|\hat Y_1(T)|^2 + |\hat Y_2(T)|^2\right)^2\;.\label{eq:ginK2}
\end{eqnarray}
\end{subequations}
Although somewhat cumbersome, it is straightforward to continue the computation for $n_Y>2$, 
where two or more singlet contractions of modular forms appear for each value of $n_Y$.

From these general results in eqs.~\eqref{eq:DiagonalKaehler} and~\eqref{eq:ginK}, one can now 
impose invariance under the $\Delta(54)$ traditional flavor symmetry to the $T'$ invariant 
contributions to the K\"ahler potential found in eqs.~\eqref{eq:Kaehlerw1} and~\eqref{eq:Kaehlerw2}. 
We see that they are compatible with the full eclectic flavor group provided that 
\begin{equation}
\kappa_1^{(0)} ~=~ \kappa_2^{(0)} \qquad\mathrm{and}\qquad \kappa^{(1)}_1 ~=~ \kappa^{(1)}_2 ~=~ \frac12\kappa^{(1)}_3\;.
\end{equation}

It is important to remark that, in contrast to the results of ref.~\cite{Chen:2019ewa}, in our setup
the $\Delta(54)$ traditional flavor symmetry prevents the appearance of non-diagonal contributions to the K\"ahler 
metric, as one can most easily read off from eq.~\eqref{eq:DiagonalKaehler}. Therefore, adding in our model an explicit dependence
on the modular forms in the K\"ahler potential does not strongly alter the phenomenological predictions obtained 
by assuming a canonical K\"ahler potential. In particular, the resulting mixing parameters of a model that includes the whole
modular dependence in $g_n(T,\bar T)$ do not differ from those described solely by the contribution proportional to 
$\kappa_1^{(0)}$.

\newpage

\subsection{Summary}
\label{sec:actionsummary}

Let us summarize our main findings of this section on the structure of the trilinear superpotential 
and bilinear K\"ahler potential of matter fields. We realize that the trilinear superpotential has 
the general structure eq.~\eqref{eq:generalW}, where the coefficients are combinations of the 
modular forms $\hat Y^{(n_Y)}_{\rep s}(T)$ detailed in table~\ref{tab:YRepresentations} with 
specific modular weights $n_Y$ and $T'$ representations $\rep s$. After discussing separately the 
constraints on the superpotential arising from $T'$ (sections~\ref{subsubsec:WwithT'} 
and~\ref{subsubsec:WwithT'andOsc}) and $\Delta(54)$ (section~\ref{subsubsec:WwithT'andDelta54}), we 
find that the twisted matter contributions to the superpotential are explicitly given by 
eq.~\eqref{eq:Z3superpotential} and eq.~\eqref{eq:Z3superpotential-withoscillators} in terms of the 
components of the matter triplet fields $\Phi_{\nicefrac{-2}{3}}=(X,Y,Z)^\mathrm{T}$ and 
$\Phi_{\nicefrac{-5}{3}}=(\tilde X,\tilde Y,\tilde Z)^\mathrm{T}$. Interestingly, the constraints 
from the symmetries reduce the number of free parameters from eleven (without traditional flavor 
symmetry) to only two (when including the traditional flavor symmetry). We then proceed to compute 
the bilinear K\"ahler potential of matter fields, assuming the most general consistent structure 
eq.~\eqref{eq:MatterKaehlerAnsatz}. We find that the restrictions arising from $T'$ and 
$\Delta(54)$ result in a diagonal K\"ahler potential, eq.~\eqref{eq:DiagonalKaehler}, implying that 
in this case nontrivial flavor mixings can only arise from the superpotential, as usually assumed. 
It should be emphasized that in these models, superpotential and K\"ahler potential transform both 
nontrivially under modular transformations, but combine to an invariant action. The eclectic nature 
of the symmetry in the TD constructions gives severe restrictions on the parameters of the theory, 
both for the superpotential and the K\"ahler potential.

\vskip 9mm
\section{Conclusions and outlook}
\label{sec:conclusions}

In the present paper we have worked out in detail a specific model that illustrates the properties 
of a new approach~\cite{Baur:2019kwi,Baur:2019iai,Nilles:2020nnc} to the flavor problem based on 
top-down (TD) model building in string theory that emphasizes the eclectic nature of the flavor 
group~\cite{Nilles:2020nnc}. The specific properties of our eclectic model are separately 
summarized in the individual sections: section~\ref{sec:spectrumsummary} reviews the representations 
including the (integer or fractional) modular weights and their nontrivial interrelations, 
section~\ref{sec:actionsummary} summarizes the power of the eclectic flavor approach to constrain 
the superpotential and the K\"ahler potential. From this construction, we derive the following 
messages for flavor model building:
\begin{itemize}
\item There is no possible scheme with just modular flavor symmetries. We always have a nontrivial 
traditional flavor group that completes the eclectic picture. This traditional flavor symmetry might 
forbid certain couplings in a given model and spoil the phenomenological predictions. The traditional 
flavor symmetry reduces the number of free parameters. A satisfactory eclectic model thus has more
predictive power than a model with just modular flavor symmetries. The interplay between the
traditional flavor group and the modular flavor symmetry is manifest in the consistency constraints
on the admissible (fractional) modular weights of matter fields.

\item One should not consider only the superpotential of the model. The K\"ahler potential plays a
crucial role as well~\cite{Chen:2019ewa}. In TD constructions, the superpotential typically
transforms nontrivially under the modular flavor symmetry. The K\"ahler potential has to compensate 
this transformation. This leads to the appearance of new free parameters that might interfere with 
the predictions derived solely from the superpotential. But again, the presence of the traditional
flavor group might reduce the number of these parameters and lead to enhanced predictive power.

\item In TD model constructions, only a subset of the possible representations and the modular
weights of the flavor group appear in the low-energy effective theory. This is true for the modular 
symmetries ($T'$ in our example) and the traditional flavor symmetry (here $\Delta(54)$) as well. 
This is a challenge for TD model building in comparison to BU-models that typically assume the presence 
of many of these possible representations. On the other hand it could lead to problems for ultraviolet 
completions of some of the BU constructions.

\item In the eclectic scheme the appearance of discrete $R$-symmetries is an unavoidable 
consequence of modular transformations. Their specific properties shall be investigated 
elsewhere~\cite{Nilles:2020tdp}.
\end{itemize}

Given these observations, one should try to intensify TD model building. Our example was motivated from
constructions based on the $\mathbbm{T}^6/\Z3\x\Z3$ orbifold~\cite{Carballo-Perez:2016ooy} and there
is a substantial landscape of heterotic orbifold models that should be explored as well. The same is 
true for models base on type II string constructions or F-theory. In fact, when we were in the final 
stage of the present paper, we became aware of ref.~\cite{Ohki:2020bpo}. This paper confirms the
eclectic picture of ref.~\cite{Nilles:2020nnc} and provides new models in the framework of magnetized 
branes in type II theories.

\section*{Acknowledgments}

We thank Alexander Baur for useful discussions on the structure of the superpotential.
The work of S.R.-S.\ was partly supported by DGAPA-PAPIIT grant IN100217, CONACyT grants F-252167 and 
278017, the Deutsche Forschungsgemeinschaft (SFB1258) and the TUM August--Wilhelm Scheer Program. The 
work of P.V. is supported by the Deutsche Forschungsgemeinschaft (SFB1258).

\newpage

\appendix

\section[T' invariant superpotential terms of T2/Z3 orbifolds]{\boldmath $T'$ invariant superpotential terms of $\mathbbm{T}^2/\Z3$ orbifolds \unboldmath}
\label{sec:Wterms}

The contributions to the trilinear superpotential of a $\mathbbm{T}^2/\Z3$ orbifold resulting from 
twisted matter fields $\Phi^i_{-\nicefrac23}=(X_i,Y_i,Z_i)^\mathrm{T}$ without oscillator 
excitations, considering only invariance under the modular symmetry $T'$ are
\begin{subequations}
\label{eq:WtermsNoOscillators}
\begin{eqnarray}
\mathcal{W}_1 & = & \frac{1}{4} \left(\hat{Y}_2(T) (4\, X_1\, X_2\, X_3 + (Y_1+Z_1) (Y_2+Z_2) (Y_3+Z_3))\right. \\
              &   & \left.-\sqrt{2} \hat{Y}_1(T) \left((Y_1+Z_1) (Y_2+Z_2)X_3 + ((Y_1+Z_1)X_2 + X_1 (Y_2+Z_2)) (Y_3+Z_3)\right)\right)\;, \nonumber \\
\mathcal{W}_2 & = & \frac{1}{4} \left(\sqrt{2} \hat{Y}_1(T) X_1 + \hat{Y}_2(T) (Y_1+Z_1)\right) (Y_2-Z_2) (Y_3-Z_3)\;,\\
\mathcal{W}_3 & = & \frac{1}{4} (Y_1-Z_1) \left(\sqrt{2} \hat{Y}_1(T) X_2 + \hat{Y}_2(T) (Y_2+Z_2)\right) (Y_3-Z_3)\;,\\
\mathcal{W}_4 & = & \frac{1}{4} (Y_1-Z_1) (Y_2-Z_2) \left(\sqrt{2} \hat{Y}_1(T) X_3 + \hat{Y}_2(T) (Y_3+Z_3)\right)\;.
\end{eqnarray}
\end{subequations}

The contributions to the trilinear superpotential arising from twisted matter fields 
$\Phi^i_{-\nicefrac53}=(\tilde X_i,\tilde Y_i,\tilde Z_i)^\mathrm{T}$ with oscillator excitations, 
considering only invariance under the modular symmetry $T'$ are
\begin{subequations}
\label{eq:WtermsOscillators}
\begin{eqnarray}
\widetilde{\mathcal{W}}_1 & = & \tfrac{1}{2\sqrt2} \hat{Y}_{\rep1'}^{(4)}(T) \left(\tilde X_2 (\tilde Y_1 + \tilde Z_1) - \tilde X_1 (\tilde Y_2 + \tilde Z_2) \right) (\tilde Y_3 - \tilde Z_3)\,,\\
\widetilde{\mathcal{W}}_2 & = & \tfrac{1}{2\sqrt2} \hat{Y}_{\rep1'}^{(4)}(T) \left(\tilde X_3 (\tilde Y_1 + \tilde Z_1) - \tilde X_1 (\tilde Y_3 + \tilde Z_3) \right) (\tilde Y_2 - \tilde Z_2)\,,\\
\widetilde{\mathcal{W}}_3 & = & \tfrac{1}{2\sqrt2} \hat{Y}_{\rep1'}^{(4)}(T) \left(\tilde X_3 (\tilde Y_2 + \tilde Z_2) - \tilde X_2 (\tilde Y_3 + \tilde Z_3) \right) (\tilde Y_1 - \tilde Z_1)\,,\\
\widetilde{\mathcal{W}}_4 & = & \tfrac{1}{2\sqrt2} \hat{Y}_{\rep1}^{(4)}(T) \,(\tilde Y_1 - \tilde Z_1) (\tilde Y_2 - \tilde Z_2) (\tilde Y_3 - \tilde Z_3)\,,\\
\widetilde{\mathcal{W}}_5 & = & \tfrac{1}{2\sqrt2} (\tilde Y_3 - \tilde Z_3) \bigg[ \tilde X_2 \left(2\,\hat Y_{\rep3,3}^{(4)}(T) \tilde X_1 + \hat Y_{\rep3,2}^{(4)}(T)(\tilde Y_1 + \tilde Z_1) \right) \\
              &   & \hspace{2.5cm} +\, (\tilde Y_2 + \tilde Z_2) \left(\hat Y_{\rep3,2}^{(4)}(T) \tilde X_1 + \hat Y_{\rep3,1}^{(4)}(T) (\tilde Y_1 + \tilde Z_1) \right)\bigg]\,, \nonumber\\
\widetilde{\mathcal{W}}_6 & = & \tfrac{1}{2\sqrt2} (\tilde Y_2 - \tilde Z_2) \bigg[ \tilde X_3 \left(2\,\hat Y_{\rep3,3}^{(4)}(T) \tilde X_1 + \hat Y_{\rep3,2}^{(4)}(T)(\tilde Y_1 + \tilde Z_1) \right) \\
              &   & \hspace{2.5cm} +\, (\tilde Y_3 + \tilde Z_3) \left(\hat Y_{\rep3,2}^{(4)}(T) \tilde X_1 + \hat Y_{\rep3,1}^{(4)}(T) (\tilde Y_1 + \tilde Z_1) \right)\bigg]\,, \nonumber\\
\widetilde{\mathcal{W}}_7 & = & \tfrac{1}{2\sqrt2} (\tilde Y_1 - \tilde Z_1) \bigg[ \tilde X_3 \left(2\,\hat Y_{\rep3,3}^{(4)}(T) \tilde X_2 + \hat Y_{\rep3,2}^{(4)}(T)(\tilde Y_2 + \tilde Z_2) \right) \\
              &   & \hspace{2.5cm} +\, (\tilde Y_3 + \tilde Z_3) \left(\hat Y_{\rep3,2}^{(4)}(T) \tilde X_2 + \hat Y_{\rep3,1}^{(4)}(T) (\tilde Y_2 + \tilde Z_2) \right)\bigg]\,, \nonumber
\end{eqnarray}
\end{subequations}
where $\hat{Y}_{\rep1}^{(4)}(T),\hat{Y}_{\rep1'}^{(4)}(T)$ and the components 
$\hat{Y}_{\rep3,j}^{(4)}(T)$, $j=1,2,3$, are given in eqs.~\eqref{eq:formsWeight4}.

\newpage


\providecommand{\bysame}{\leavevmode\hbox to3em{\hrulefill}\thinspace}

\end{document}